\documentclass{aa}
\usepackage{color}
\usepackage{graphicx}
\usepackage{txfonts}
\begin{document}

   \title{Highly collimated microquasar jets as efficient cosmic-ray sources}

   \subtitle{}

   \author{G.J. Escobar
          \inst{1,3}
          \and
          L.J. Pellizza\inst{2}
          \and
          G.E. Romero\inst{1}
          } %\fnmsep\thanks{Just to show the usage
          %of the elements in the author field}

   \institute{Instituto Argentino de Radioastronom\'ia, CONICET-CICBA, Argentina \and Instituto de Astronom\'ia y F\'isica del Espacio (CONICET-UBA), C.C. 67, Suc. 28, C1428ZAA, Buenos Aires, Argentina \and Physics and Astronomy Department Galileo Galilei, University of Padova, Vicolo dell’Osservatorio 3, I–35122, Padova, Italy\\
              \email{gescobar@iar.unlp.edu.ar // gastonjavier.escobar@unipd.it}}

   \date{}

% \abstract{}{}{}{}{} 
% 5 {} token are mandatory
 
  \abstract
   {Supernova remnants are currently believed to be the main sites where Galactic cosmic rays originate. This scenario, however, fails to explain some of the features observed in the cosmic-ray spectrum. Microquasars have been proposed as additional candidates, because their non-thermal emission indicates the existence of efficient particle acceleration mechanisms in their jets. Only a few partial attempts have been made so far to quantify the contribution of microquasars to the Galactic cosmic-ray population. A promising scenario envisages the production of relativistic neutrons in the jets, that decay outside the system injecting relativistic protons to the surroundings. The first investigations of this scenario suggest that microquasars might be fairly alternative cosmic-ray sources.}
   {We aim at assessing the role played by the degree of collimation of the jet on the cosmic-ray energetics in the neutron-carrier scenario, as well as the location and size of the emission region, and the interactions of protons with photon fields. Our goals are to explain the Galactic component of the observed proton cosmic-ray spectrum at energies higher than $\sim 10~\mathrm{GeV}$ and to relate the mentioned jet properties with the power and spectral index of the produced cosmic rays.}
   {We improve previous analytical models of relativistic particle transport in microquasar jets by including prescriptions for the jet geometry and convection within it. We introduce the neutron component through catastrophic terms that couple the proton and neutron transport equations, and then compute the escape and decay of these neutrons. Finally, we follow the propagation of the decay products and obtain the proton cosmic-ray spectrum once the particles reach the interstellar medium.}
   {We find that collimated jets, with compact acceleration regions close to the jet base, are very efficient sources that could deliver a fraction of up to $\sim 0.01$ of their relativistic proton luminosity into cosmic rays. Collimation is the most significant feature regarding efficiency; a well collimated jet might be $\sim 4$ orders of magnitude more efficient than a poorly collimated one. These sources produce a steep spectral index of $\sim 2.3$ at energies up to $\sim 10~\mathrm{TeV}$.}
   {Single microquasars may rival supernova remnants regarding the power injected to the interstellar medium in cosmic rays. The main advantage of the former is the production of a steeper spectrum than the latter, closer to what is observed. The predictions of our model may be used to infer the total contribution of the population of Galactic microquasars to the cosmic ray population, and therefore to quantitatively assess their significance as cosmic-ray sources.}


   \maketitle
%
%-------------------------------------------------------------------

\section{Introduction}

The origin of Galactic cosmic rays (CRs) is still a matter of debate. At present, supernova remnants (SNRs) are considered the most plausible sites where these particles accelerate up to high energies, diffusing afterwards through the ambient medium. Even though this paradigm explains the CR power observed at Earth, it fails to reproduce simultaneously the main characteristics observed in the CR spectrum, such as the energy at which the transition to extragalactic CRs is found, the slope of the spectrum, or the observed anisotropy and chemical composition (\citealt{Ginzburg1963}; see also \citealt{Blasi2013} for a recent review). This lack of agreement may reflect the existence of alternative sources of CRs, but also a poor knowledge of either the dynamics of CRs or the properties of the ambient medium through which these particles propagate. 

Diverse attempts to explain the observed CR spectrum invoking contributions of different sources of particle acceleration have been made. Proposals include that CRs originate in %the jets of active galactic nuclei \citep[AGNs;][]{Biermann1994, Dutan2015, Mbarek2019, Rodrigues2021}, 
pulsars \citep[e.g.][]{Bednarek2002, Bednarek2004}, microquasars \citep[MQs;][]{Hillas1984, Heinz2002, Bednarek2005}, winds of massive stars or stellar clusters, and enhanced star-forming regions \citep[e.g.][]{Morlino2021, Peretti2021}. \citet{RomeroVila2009} also suggest that MQs might be responsible for the positron component of the CR spectrum. These scenarios are based on the presence of outflows developing strong magnetized shocks in which particles accelerate to relativistic energies. The same magnetic fields are, however, also efficient in trapping and cooling particles, therefore reducing the power and maximum particle energies achieved.

Recently, we have developed a different scenario \citep{Escobar2021} in which the escape of CRs from magnetically confined jets of MQs is mediated by relativistic neutrons. The presence of charged particles accelerated to relativistic energies in MQ jets follows from observations of the non-thermal radio and gamma-ray emission of these systems. Observations show that hadronic content is present at least in some MQs \citep[][]{Migliari2002, DiazTrigo2013}, which is also supported by theoretical models of jet launching mechanisms \citep{Blandford&Payne1982}.
The production of relativistic neutrons is an inescapable consequence of these facts, given that a fraction of the interactions between relativistic and thermal protons produce neutrons. These neutrons carry most of the energy of the progenitor projectiles and escape freely from magnetic fields that confine the jet. Indeed, relativistic neutron production up to $\sim$EeV energies has been explored in the jets of AGNs \citep[e.g.][]{Sikora1989, Atoyan2003}, although the large spatial scales of these systems prevent that neutrons escape before decaying. Our model predicts that MQ jets with large kinetic luminosities and low velocities produce neutrons up to PeV energies that decay into protons and electrons outside the system, therefore directly injecting CRs in the surrounding medium \citep{Escobar2021}. These CRs display steep spectra (i.e. spectral index $\gtrsim 2.5$), and may carry up to $\sim 10^{48-49}\ \mathrm{erg}$ throughout the whole MQ life, only $\sim 1-2$ orders of magnitude less than a SNR.

The scenario of \citet{Escobar2021} is based on the MQ jet model of \citet{RomeroVila2008}. Its main ingredient is inelastic $p$-$p$ scattering, which produces relativistic neutrons from the abundant relativistic protons accelerated in jet internal shocks. This process competes mainly with proton adiabatic cooling (except at sites near the jet base and very high energies at which proton synchrotron dominates cooling), assumed to operate in a conic jet. Some features of the model are adopted from the observed properties of one of the best studied MQs, namely \object{Cygnus X-1} \citep{Pepe2015}. Recently, \citet{Kantzas2021} have developed a different model for the same MQ, which includes some ingredients not present in that of \citet{RomeroVila2008}. The first one is a parabolic geometry for the jet, which directly affects the adiabatic cooling regulating losses, and should therefore affect neutron production. Also, convection within the jet is treated in the model of \citet{Kantzas2021}. This may transport relativistic protons to lower density regions, decreasing the $p$-$p$ scattering rate. Finally, second order processes, such as photomeson production in the synchrotron radiation field of the jet, have been shown to contribute to $\gamma$-ray emission by these authors, and therefore may contribute also to the neutron production. These processes were absent in the work of \citet{Escobar2021} in which only estimates of the contribution from external photon fields were implemented. In this work we improve upon the model of these authors, in order to explore the effects of the jet geometry, convection and $p$-$\gamma$ interactions on CR production mediated by relativistic neutrons.

 This paper is organised as follows. In Sect.~\ref{Sec:JetModel} we describe the jet model and the neutron production mechanisms. In Sect.~\ref{Sec:Results} we explore the neutron and cosmic-ray luminosities and spectra for different scenarios in our model. We discuss the results and present our conclusions in Sect.~\ref{Sec:Conclusions}.

\section{Jet model}
\label{Sec:JetModel}

\begin{figure}
    \centering
    \includegraphics[width=0.37\hsize]{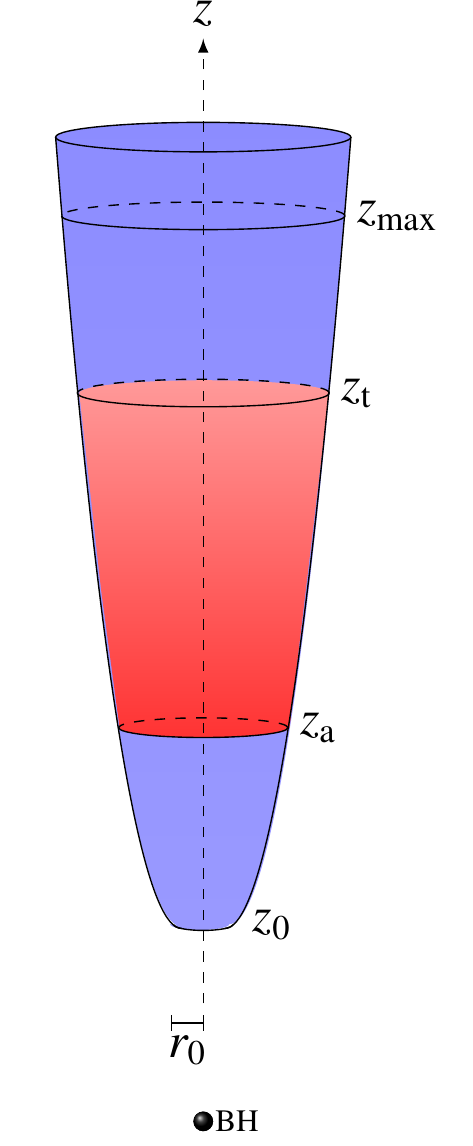}
    \caption{Schematic picture of the jet (not to scale), showing a jet launched at a distance $z_{0}$ from the compact object. The acceleration region, shaded in red, is defined between $z_{\mathrm{a}}$ and $z_{\mathrm{t}}$, while the emission region spans over the range $z_{\mathrm{a}} \leq z \leq z_{\mathrm{max}}$.}
    \label{Fig:Jet}
\end{figure}

We use a MQ jet model based on that described in \citet{Escobar2021}. We refer the reader to this work for details of the original model; here we stress on the new developments. We consider a lepto-hadronic jet of luminosity $L_{\mathrm{jet}}$ and bulk velocity $v_{\mathrm{jet}}$, corresponding to a bulk Lorentz factor $\Gamma$. The jet is launched at a distance $z_{0}$ from the compact object (see Fig. \ref{Fig:Jet}). The magnetic field, $B$, is computed assuming equipartition between kinetic and magnetic energies at the jet base ($z_{0}$), and then adopting a power-law function of the distance along the jet, $B\propto z^{-m}$, $m$ being the magnetic index. A fraction of the jet power is deposited in relativistic particles once the flow develops an acceleration mechanism. The distance at which this acceleration region begins is denoted by $z_{\mathrm{a}}$, and its top by $z_{\mathrm{t}}$, while the emission region extends up to $z_{\mathrm{max}} \geq z_{\mathrm{t}}$. The shape of the emission region is parameterised by

\begin{eqnarray}
r(z) = r_{0} \left( \frac{z}{z_{0}}\right)^{\alpha},
\end{eqnarray}

\noindent where $r_{0}$ is the jet radius at its base and $0 < \alpha \leq 1$ describes the jet geometry. The jet collimation increases with decreasing $\alpha$; the values $\alpha =1/2$ and $\alpha = 1$ represent parabolic and conic shapes, respectively, and the jet becomes almost cylindrical as $\alpha \to 0$. This prescription allows us to vary the strength of adiabatic cooling, which competes with the $p$-$p$ interactions that produce neutrons.

We assume that the relativistic population takes 10\% of the jet luminosity \textbf{($q_\mathrm{rel}=0.1$)}. This fraction is distributed between protons and electrons according to a hadron-to-lepton power-ratio $a$. The discussion about the acceleration mechanism is beyond the scope of this work; here we just mention that particles accelerated up to relativistic energies in astrophysical sources are likely to develop a power-law distribution once a steady state is achieved \citep[e.g.][]{Schlickeiser2002}. Therefore, assuming that the acceleration time scale is much shorter than those of other processes, we parameterise the injection as a power law in the proton energy, $E_{\mathrm{p}}$, with spectral index $p$ and an exponential cutoff at energy $E^{\mathrm{max}}_{\mathrm{p}}(z)$, 

\begin{eqnarray}
Q_{\mathrm{p}}(E,z) = Q_{0} E^{-p}_{\mathrm{p}} \exp{\left(-\frac{E_{\mathrm{p}}}{E^{\mathrm{max}}_{\mathrm{p}}(z)}\right)}~f(z).
\end{eqnarray}

\noindent The maximum energy $E^{\mathrm{max}}_{\mathrm{p}}$ is achieved when the total loss rate equals the acceleration rate of the particles (see discussion below), and $Q_{0}$ is obtained by normalising the injection function to the total power of the relativistic proton population. The function $f(z)$ is chosen to represent the result of the convective motion that produces particle injection beyond the acceleration region. Therefore, we choose $f(z) = 1$ for $z_\mathrm{a} \leq z \leq z_{\mathrm{t}}$ and a power-law $f(z) = \left(z/z_{\mathrm{t}}\right)^{-\delta}$ for $z > z_\mathrm{t}$. The index $\delta > 0$ is left free to vary the convective strength (lower values representing stronger convection; see discussion in Sect. \ref{Sec:Results}).

The steady-state solution for the densities of the relativistic hadron populations is given by the coupled transport equations

\begin{eqnarray}
\frac{\partial}{\partial E}[b_{\mathrm{p}}N_{\mathrm{p}}] + t^{-1}_{\mathrm{esc}} N_{\mathrm{p}} = Q_{\mathrm{p}} - \Lambda_{\mathrm{p}},\\
t^{-1}_{\mathrm{esc,n}} N_{\mathrm{n}} = Q_{\mathrm{n}}.
\label{transp}
\end{eqnarray}

\noindent Here $N_{\mathrm{p (n)}}$ is the proton (neutron) steady-state particle density, $t_{\mathrm{esc}}$ and $t_{\mathrm{esc,n}}$ are the proton and neutron escape rates, respectively, and $b_{\mathrm{p}}$ accounts for the total energy loss rate for protons. The terms $Q_{\mathrm{n}}$ and $\Lambda_{\mathrm{p}}$ represent the source rate of neutrons and the related sink rate of protons, respectively. All quantities depend on the energy of the corresponding particle population and the distance $z$. The boundary condition is $N_{\mathrm{p}}(E_\mathrm{p}^{\mathrm{max}}(z), z) = 0$. The effect of any term depending on the derivatives with respect to $z$ is included effectively by the function $f(z)$, and therefore not included explicitly in Eq.~\ref{transp}.

Proton-proton inelastic collisions are a source mechanism for relativistic neutrons through the branching channels in which the relativistic proton is destroyed, yielding a neutron that carries approximately half of the proton energy plus several secondary particles. The total cooling rate is

\begin{eqnarray}
t^{-1}_{\mathrm{pp}} = \sigma_{\mathrm{pp}} n_{\mathrm{p}} v_{\mathrm{rel}} K_{\mathrm{pp}},
\end{eqnarray}

\noindent where $\sigma_{\mathrm{pp}}$ and $K_{\mathrm{pp}}$ are the cross section and inelasticity of the process, $v_{\mathrm{rel}}$ is the relative velocity of colliding particles, which we take as $c$, and $n_{\mathrm{p}} = n_\mathrm{b} + n_\mathrm{w}$ is the target proton density, which accounts for the contribution of protons from the jet bulk material ($n_\mathrm{b}$) and the wind of the companion star ($n_\mathrm{w}$). These densities are given by

\begin{eqnarray}
n_\mathrm{b} &=& \frac{L_\mathrm{jet} (1 - q_\mathrm{jet})}{m_\mathrm{p}c^{2}v_\mathrm{jet}\pi r(z)^{2}},\\
n_\mathrm{w} &=& \frac{\Gamma \dot{M}}{4\pi (d^{2} + z^2) v_\mathrm{w}m_\mathrm{H}},
\end{eqnarray}

\noindent in the jet reference frame. In the latter equation, $\dot{M}$ is the mass-loss rate of the companion, $v_\mathrm{w}$ its wind velocity, $d$ its distance to the compact object, and $m_\mathrm{H}$ is the mass of the hydrogen atom. We adopt the standard velocity profile for a line-driven wind given by \citet[][]{Lamers1999}. A discussion about the impact of the wind density on the model and its results can be found in \citet[][]{Escobar2021}. Both contributions are treated separately to assess their individual significances. We adopt the parameterisation of the cross section given by \citet{Kafexhiu2014}. Regarding the neutron channel, we adopt a conservative branching ratio of 0.16 \citep[see discussion in Sect.~2 of][]{Escobar2021}. Thus, the proton sink rate due to this process is given by $\Lambda_{\mathrm{pp}}(E_{\mathrm{p}}) = 0.16 t^{-1}_{\mathrm{pp}}(E_{\mathrm{p}})N_{\mathrm{p}}(E_{\mathrm{p}})$, and the corresponding neutron source rate is $Q_{\mathrm{n}}^{\mathrm{(pp)}}(E_{\mathrm{n}})$ = $\Lambda_{\mathrm{pp}}(E_{\mathrm{p}})$, with $E_{\mathrm{n}} = 0.5 E_{\mathrm{p}}$.

Some channels of photohadronic interactions also produce relativistic neutrons. We consider the inelastic collisions between protons and radiation fields, both internal and external to the jet. To compute the cooling rate, inelasticities, and inclusive cross sections we adopt the prescription given in \citet{Atoyan2003}. For the single-pion channel (photon energies $\varepsilon$ between $200$~MeV and $500$~MeV, in the reference frame of the proton) the cross section can be approximated as $\sigma_{1} \approx 340~\mu$barn and the inelasticity as $K_{1} = 0.2$. In the multi-pion channel ($\varepsilon \geq 500$~MeV) these values are $\sigma_{2} \approx 120~\mu$barn and $K_{2} \approx 0.6$. For both channels, the probability of conversion of a proton to a neutron is $\approx 0.5$. The cooling rate is given by

\begin{eqnarray}
t^{-1}_{\mathrm{p\gamma}}(E_{\mathrm{p}}) = \int_{\frac{\varepsilon_{\mathrm{th}}}{2\gamma_{\mathrm{p}}}}^{\infty}\mathrm{d}\varepsilon \,\frac{n_{\mathrm{ph}}c}{2\varepsilon^{2}\gamma_{\mathrm{p}}^{2}} \int_{\varepsilon_{\mathrm{th}}}^{2\varepsilon \gamma_{\mathrm{p}}}\varepsilon'\sigma_{\mathrm{p\gamma}}(\varepsilon')K_{\mathrm{p\gamma}}(\varepsilon') \mathrm{d}\varepsilon',  \end{eqnarray}

\noindent where $\gamma_{\mathrm{p}}$ is the proton Lorentz factor, $\varepsilon$ the target photon energy, $\varepsilon_{\mathrm{th}}$ the energy threshold, $n_{\mathrm{ph}}$ the differential number density of target photons, and $\sigma_{\mathrm{p\gamma}}$ and $K_{\mathrm{p\gamma}}$ are the cross section and inelasticity functions, respectively. Similarly, the collision rate is given by

\begin{eqnarray}
\nu_{\mathrm{p\gamma}}(E_{\mathrm{p}}) = \int_{\frac{\varepsilon_{\mathrm{th}}}{2\gamma_{\mathrm{p}}}}^{\infty}\mathrm{d}\varepsilon \,\frac{n_{\mathrm{ph}}c}{2\varepsilon^{2}\gamma_{\mathrm{p}}^{2}} \int_{\varepsilon_{\mathrm{th}}}^{2\varepsilon \gamma_{\mathrm{p}}}\varepsilon'\sigma_{\mathrm{p\gamma}}(\varepsilon') \mathrm{d}\varepsilon'.  \end{eqnarray}

\noindent The proton and neutron energies are related by $E_{\mathrm{n}} = (1 - \bar{K}_{\mathrm{p}\gamma})E_{\mathrm{p}}$, where $\bar{K}_{\mathrm{p}\gamma} = \nu_{\mathrm{p\gamma}}^{-1}t^{-1}_{\mathrm{p\gamma}}$ is the mean inelasticity. The proton sink rate of this interaction is given by $\Lambda_{\mathrm{p}\gamma} = 0.5 t^{-1}_{\mathrm{p\gamma}}N_{\mathrm{p}}$, while the neutron source rate is $Q_{\mathrm{n}}^{\mathrm{(p\gamma)}}(E_{\mathrm{n}}) = \Lambda_{\mathrm{p\gamma}}(E_{\mathrm{p}})$. Finally, the total proton sink rate and total neutron source rate are given by $\Lambda_{\mathrm{p}} = \Lambda_{\mathrm{pp}} + \Lambda_{\mathrm{p}\gamma}$ and $Q_{\mathrm{n}} = Q_{\mathrm{n}}^{\mathrm{(pp)}} + Q_{\mathrm{n}}^{\mathrm{(p\gamma)}}$, respectively.

The relativistic populations also suffer non-radiative losses, such as escape, decays, and adiabatic work. We estimate the escape rate of protons as $t^{-1}_{\mathrm{esc}} = v_{\mathrm{jet}}(z_{\mathrm{
max}} - z)^{-1}$, representing the loss through the head of the jet. The fraction of particles that escape sideways is of the order of $r_\mathrm{g} / r(z) \ll 1$, with $r_\mathrm{g}$ the gyroradius, so we neglect it. The previous condition is not fulfilled by particles with energies close to the maximum energy; the power carried by these particles, however, is negligible with respect to the total power of the population. We model the escape rate of neutrons as $t^{-1}_{\mathrm{esc,n}} = c(z_{\mathrm{max}} - z)^{-1}$. As disussed in \citet{Escobar2021}, this is a conservative estimate, even though our main results do not depend on the exact value of such quantity. In addition, we neglect neutron decay within the jet due to the small size of the latter compared to the mean decay length at any relevant energy. On the other hand, the adiabatic loss-rate for protons is given by

\begin{eqnarray}
t^{-1}_{\mathrm{ad}} = \frac{2\alpha}{3z}v_{\mathrm{jet}},
\label{Ec:adiabatic}
\end{eqnarray}

\noindent given that these particles are relativistic.

The acceleration rate can be written as 

\begin{eqnarray}
t_{\mathrm{acc}}^{-1} = \frac{\eta c e B}{E},
\end{eqnarray}

\noindent where $\eta$ is an efficiency parameter and $e$ is the elementary charge. Protons also cool through radiative losses, including the already mentioned inelastic collisions with matter and radiation, and synchrotron emission. The formulae for the synchrotron losses can be found in \citet{Blumenthal1970}. Neutrons suffer neither adiabatic nor radiative losses.

\section{Neutron and cosmic-ray production}
\label{Sec:Results}

 In this section we investigate the production of relativistic neutrons for several scenarios comprising different features of the emission region, the convection of particles, the collimation of the jet, and the presence of different photon fields. %Neutrons are produced via collisions of relativistic protons with matter.
Table \ref{Tab:modelparameters} shows the parameters for the fiducial case, which is nearly the same used in \citet{Escobar2021}, except for the minimum energy and acceleration efficiency, whose values were adopted to ensure that in all cases acceleration overtakes losses. This scenario has been adapted from \citet{Pepe2015} to represent the jet of \object{Cygnus X-1}. Regarding the features under investigation, this is a conic jet with a compact acceleration region near its base, which is also the emission region (i.e., convection is negligible). No photon fields are taken into account in this case; we include the contribution of photohadronic interactions separately in Sect.~\ref{Sec:photohadronic}. 

\begin{table}[ht!]
	\caption{Jet parameters for the fiducial model.}
	\centering
	\begin{tabular}{llrl}
	\hline
	\hline
	Parameter & Symbol & Value & Units\\
	\hline
	Jet luminosity & $L_{\mathrm{jet}}$ & $10^{38}$ & $\mathrm{erg\,s^{-1}}$ \\
	Jet-base half-opening angle$^{\dagger}$ & $\theta_{\mathrm{jet}}$ & $2$ & ${\mathrm{deg}}$\\
	Bulk Lorentz factor & $\Gamma$ & $1.25$ &\\
	Relativistic power fraction & $q_{\mathrm{rel}}$ & $0.1$ &\\
	Proton-electron power ratio & $a$ & $39$ &\\
	Jet launching distance & $z_{0}$ & $1.1\times10^{8}$ & $\mathrm{cm}$\\
	Base of acceleration region & $z_{\mathrm{a}}$ & $2.8\times10^{8}$ & $\mathrm{cm}$\\
	Top of acceleration region & $z_{\mathrm{t}}$ & $1.9\times10^{12} $ & $\mathrm{cm}$\\
	End of emission region$^{\ddag}$ & $z_{\mathrm{max}}$ & $1.9\times10^{12}$ & cm\\
	Acceleration efficiency & $\eta$ & $1.5 \times 10^{-2}$ & \\
	Magnetic power-law index & $m$ & $1.9$ &\\
	Injection spectral index & $p$ & $2.4$ &\\
	Geometric index & $\alpha$ & 1 &\\
	Convection index & $\delta$ & - &\\
	\hline
	\end{tabular}
	\tablefoot{$^{\dagger}\, \tan\theta_{\mathrm{jet}} = r_{0} / z_{0}$.
	$^{\ddag}$\, In the fiducial case we consider the emission region as the same as the acceleration one.}
    \label{Tab:modelparameters}
	\end{table}

\begin{figure*}[ht]
    \centering
    \includegraphics[width=0.95\hsize]{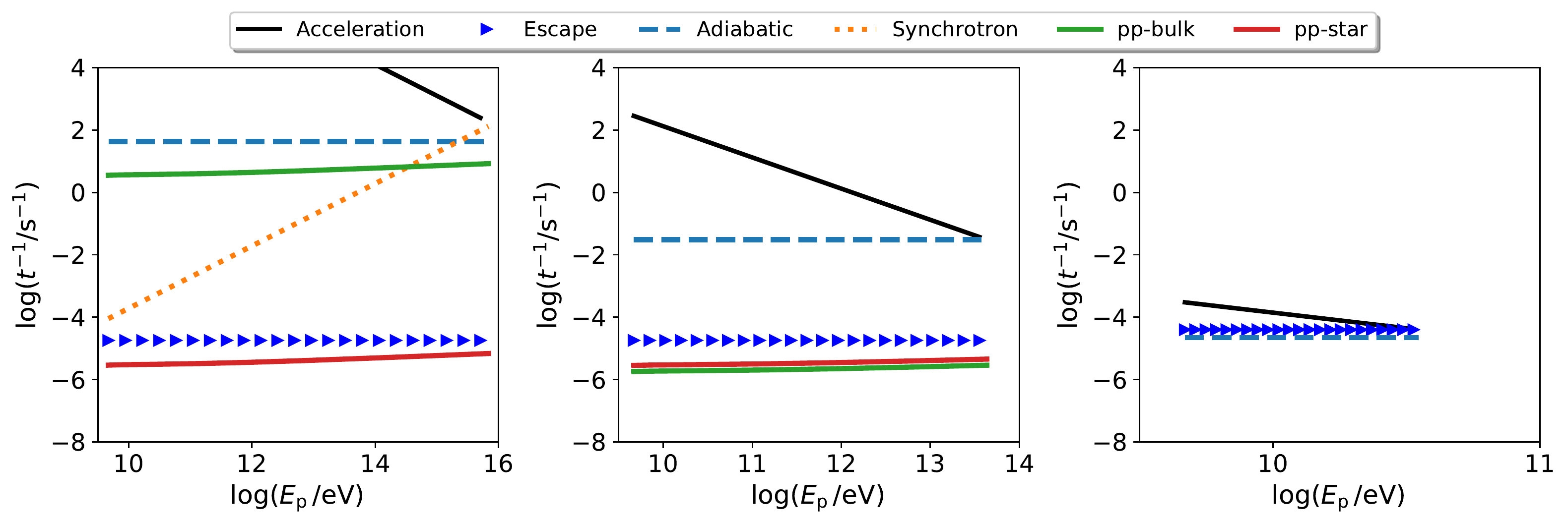}
    \caption{Proton cooling rates for the case $z_{\mathrm{t}}=z_{\mathrm{max}} = 10^{15}$~cm, at the base (left panel), logarithmic midpoint (middle panel), and the end of the acceleration region (right panel).}
    \label{Fig:lossrates}
\end{figure*}

\subsection{Acceleration region and convection}
\label{Subsec:accel&conv}

We first explore several models with different sizes for the acceleration region, starting from a compact one with $z_{\mathrm{t}}= 1.9\times 10^{12}$~cm, and extending it up to $z_{\mathrm{t}} = 10^{15}$~cm. The last value is the extension of the jet of \object{Cygnus X-1} reported by \citet{Stirling2001}, showing that particle acceleration might develop at much higher distances than those reported by \citet{Pepe2015} \citep[cf.][]{Kantzas2021}. In Fig. \ref{Fig:lossrates} we show the proton loss and acceleration rates at different sites of the jet for the model with the most extended acceleration region. It is important to stress that, except for escape, energy losses at a given position $z$ in the jet do not vary with the size of the emission region, therefore the loss curves of Fig.~\ref{Fig:lossrates} apply to any of the aforementioned models.

We observe that inelastic $p$-$p$ losses, responsible for the production of relativistic neutrons, are larger close to the base of the jet and decrease upwards. This is due to the decreasing density of bulk protons as the jet expands. The most important process that competes against $p$-$p$ is adiabatic energy loss, except for the highest energies at the jet base, where synchrotron radiation dominates, and for all energies at the jet top, where the escape losses slightly exceed adiabatic ones. Adiabatic loss and $p$-$p$ scattering are of the same order of magnitude at the jet base, but upwards $p$-$p$ losses decrease faster than adiabatic work, becoming orders of magnitude smaller. These features suggest that conditions for neutron production are most favourable at the base. 

\begin{figure}
    \centering
    \includegraphics[width=0.99\hsize]{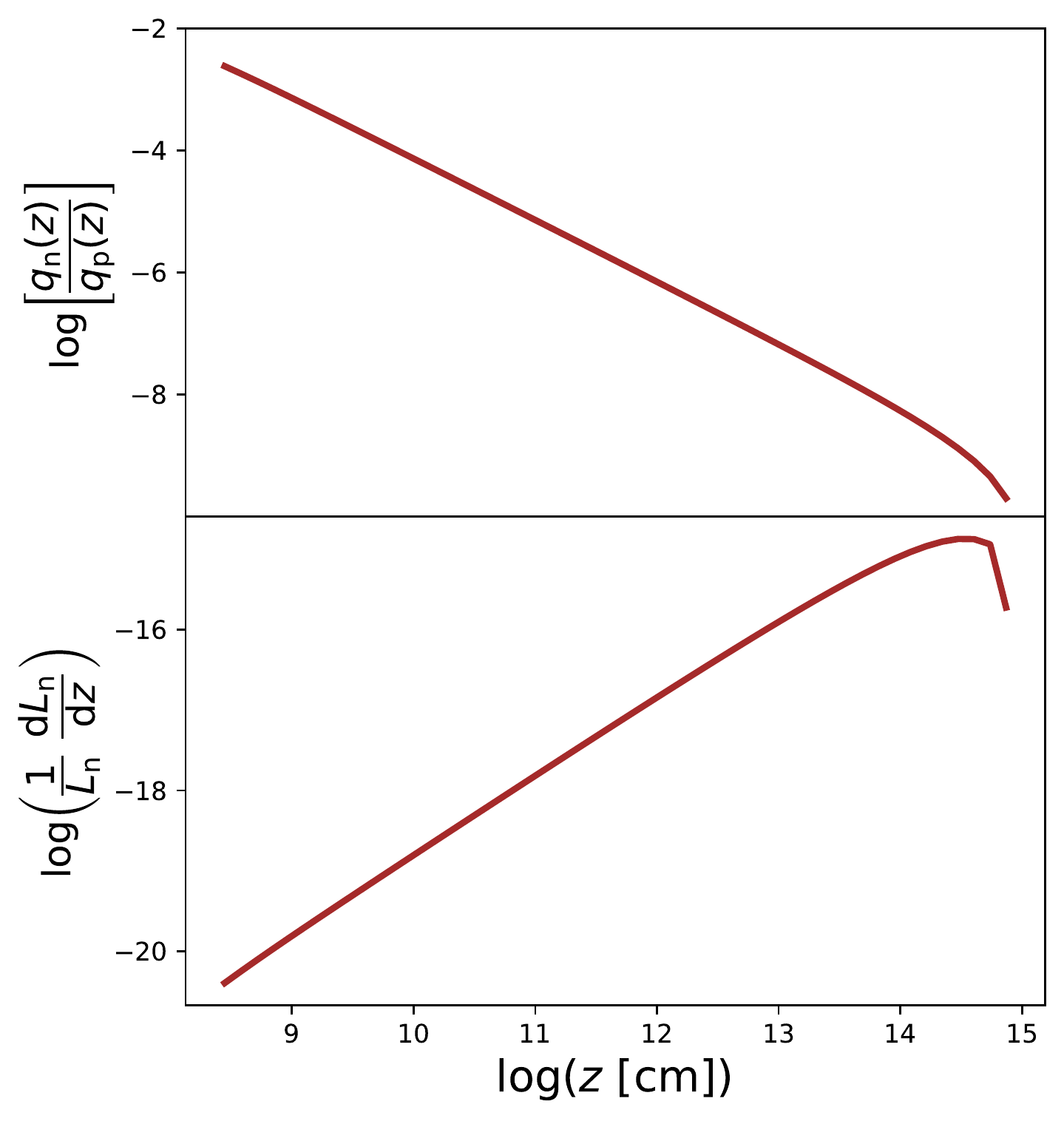}
    \caption{\emph{Top}. Ratio of neutron emissivity to proton volumetric injection rate. \emph{Bottom}. Neutron luminosity distribution normalised with total luminosity $L_{\mathrm{n}}$}
    \label{Fig:neutrontoproton}
\end{figure}

Fig.~\ref{Fig:neutrontoproton} supports the interpretation expressed above, showing that the fraction of power per unit volume injected relativistic protons, $q_{\mathrm{p}}$, that is transferred to neutron emissivity, $q_{\mathrm{n}}$, decreases with $z$, where

\begin{eqnarray}
q_{\mathrm{p(n)}}(z) = \int E_{\mathrm{p(n)}}Q_{\mathrm{p(n)}}(E_{\mathrm{p(n)}},z)\,\mathrm{d}E_{\mathrm{p(n)}}.
\label{Ec:emissivities}
\end{eqnarray}

\begin{figure}
    \centering
    \includegraphics[width=0.95\hsize]{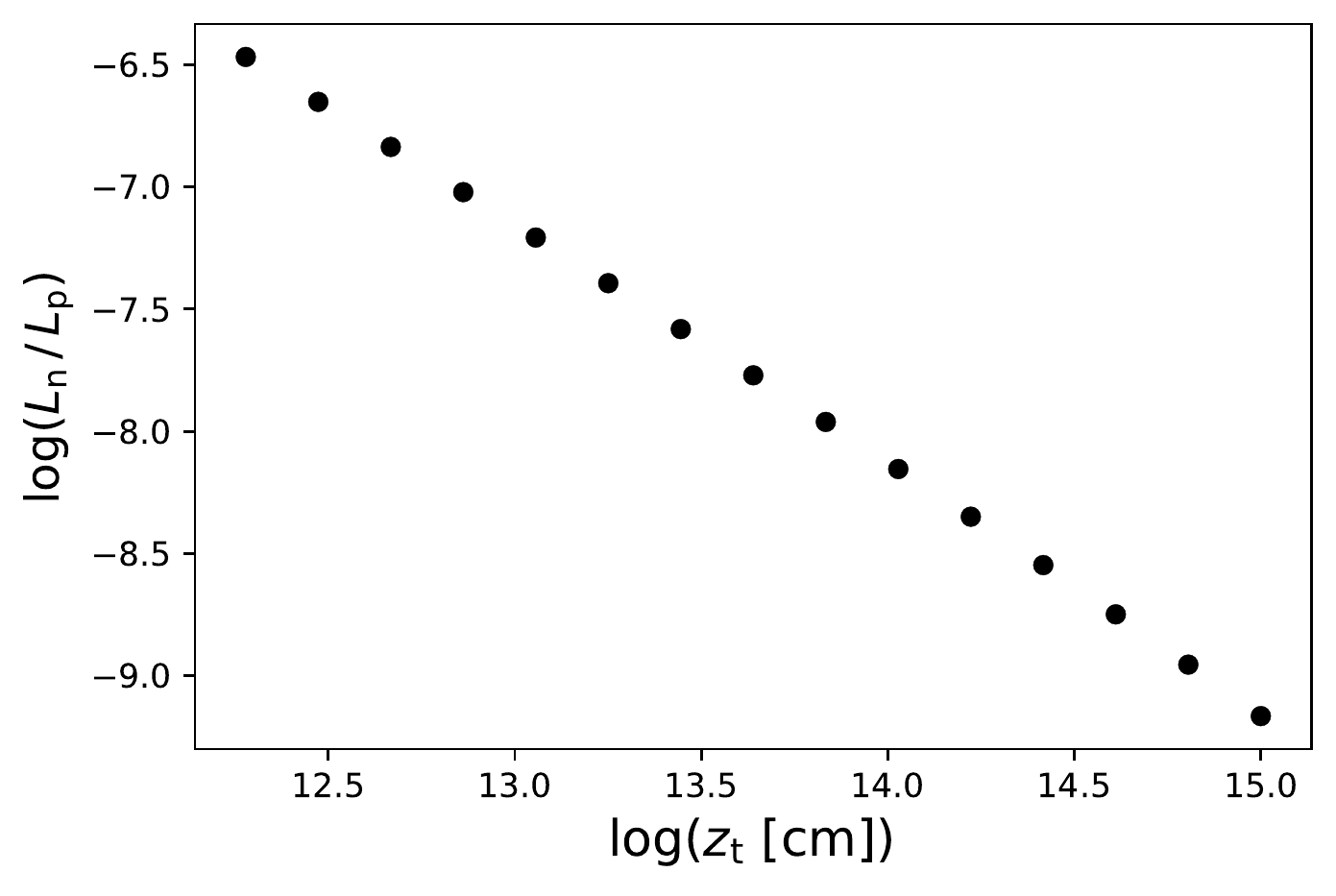}
    \caption{Fraction of the luminosity of relativistic protons transferred to neutrons, as a function of the size of the acceleration region, defined by the position of its top $z_\mathrm{t}$. Here, $L_{\mathrm{p}}$ stands for the total luminosity in the proton relativistic population.}
    \label{Fig:efficiencyvsztop}
\end{figure}

\noindent Despite the fact that the jet base is more efficient in transferring energy to the neutron channel, the total neutron luminosity, $L_{\mathrm{n}}$, is still dominated by the jet top due to its far larger volume. At the same total power injected into relativistic particles, these facts render extended acceleration regions less efficient than compact ones for neutron production. This is because in compact acceleration regions most of the power is released near the jet base, where neutrons can be efficiently produced, whereas extended ones inject most of their energy in large regions far from the base, where neutron production is not efficient. Fig.~\ref{Fig:efficiencyvsztop} shows that indeed the fraction of the total luminosity of relativistic particles transferred into neutrons decreases with increasing acceleration region size. The linear trend with $\log z_{\mathrm{t}}$ arises from the fact that for a conic jet adiabatic losses dominate the proton cooling (no dependence with $z$), while for neutrons the dominant loss process is escape (varies according to $\sim z^{-1}$). Therefore, jets with compact acceleration regions near their base are the best candidates for injecting neutrons into their surroundings.

It is important to recall that this behaviour is driven by the competition between $p$-$p$ scattering and adiabatic losses. The latter are directly related to the jet degree of collimation (see Eq. \ref{Ec:adiabatic}), and therefore we expect that collimation plays a major role in neutron production. We defer this discussion to Sect.~\ref{Subsect:collimation} 

\begin{figure}
    \centering
    \includegraphics[width=0.99\hsize]{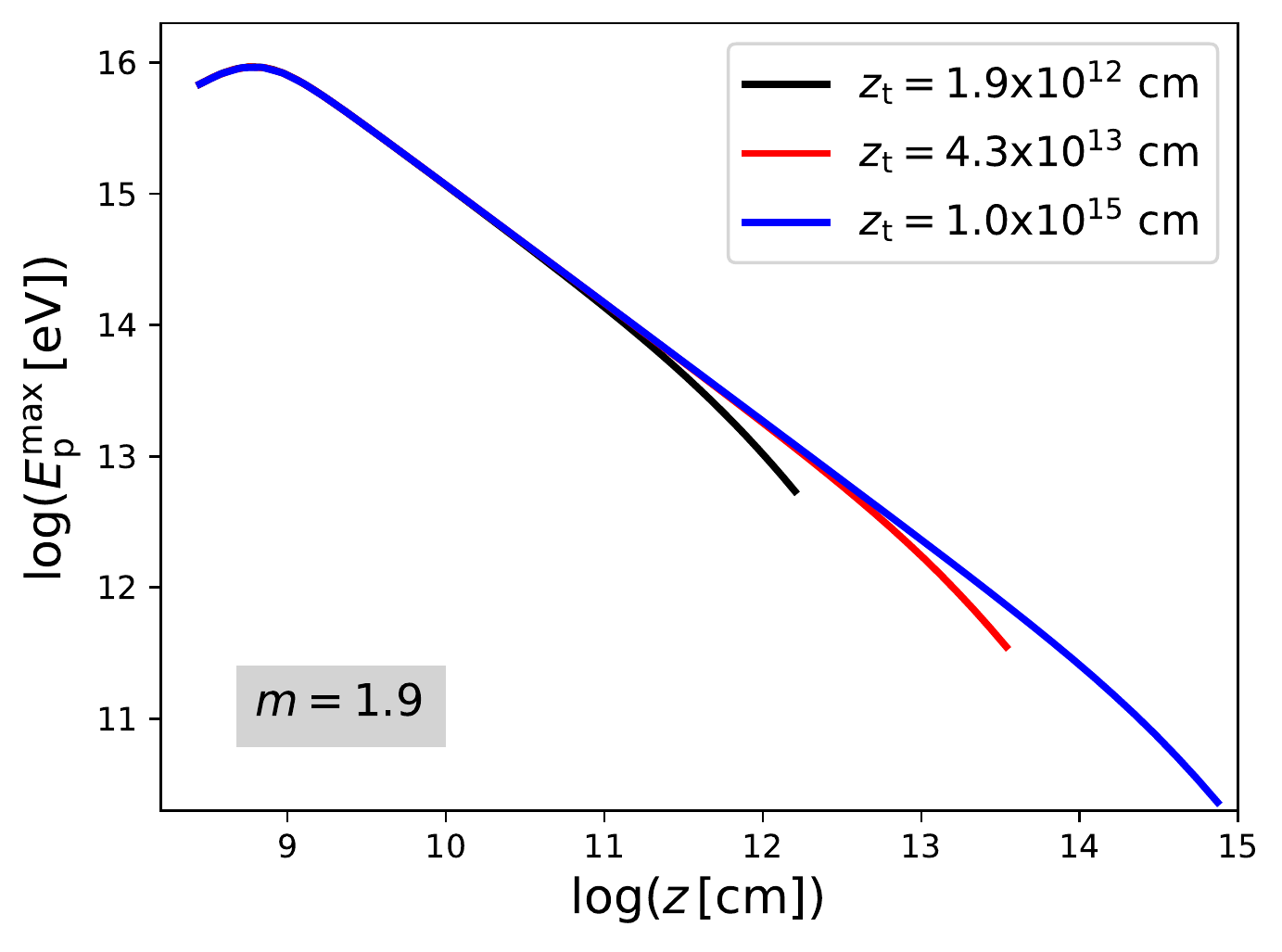}
    \includegraphics[width=0.99\hsize]{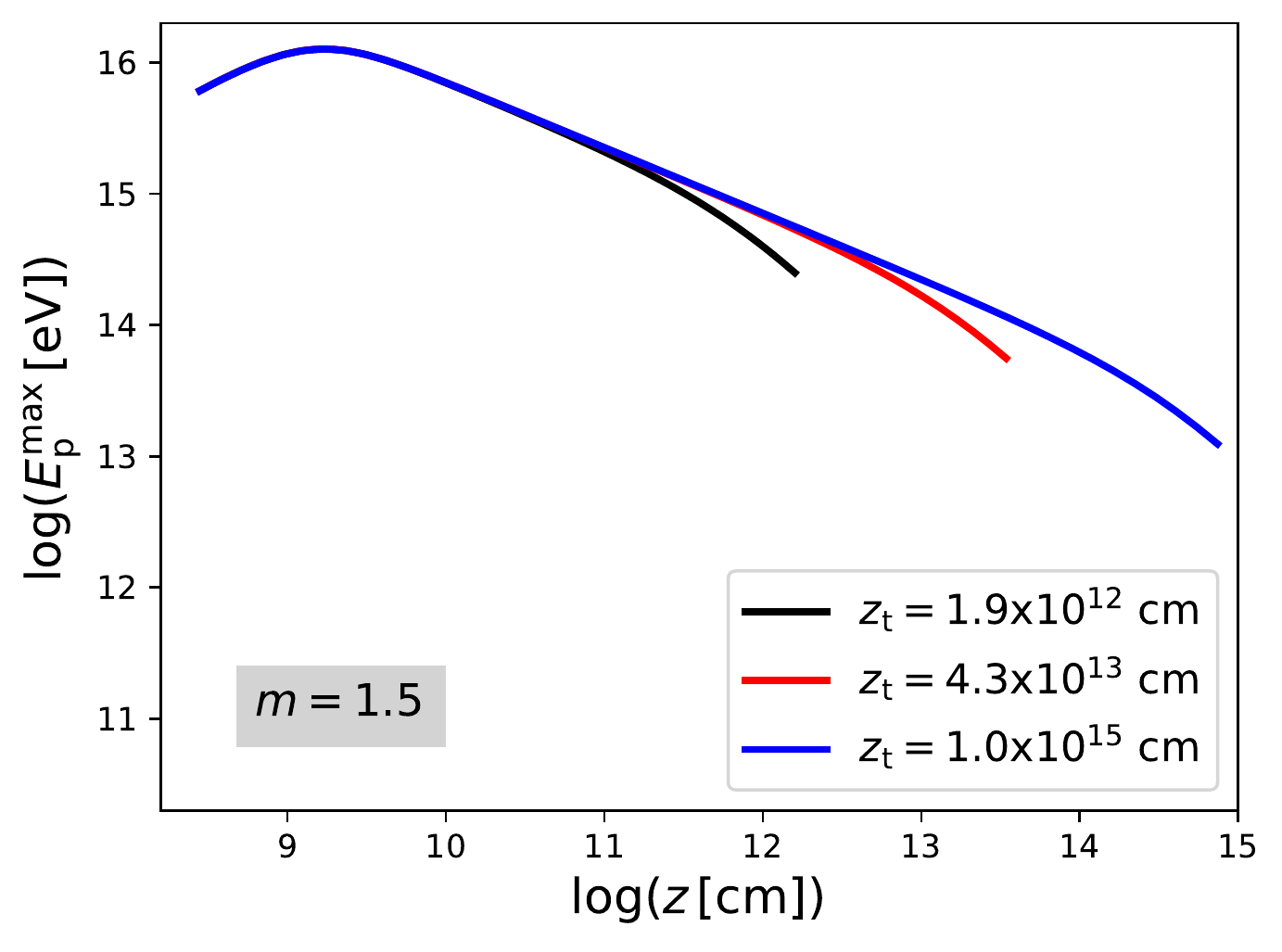}
    \caption{Maximum energy of protons as a function of distance $z$ along the jet for models comprising three different values of the acceleration region top $z_\mathrm{t}$. The top (bottom) panel corresponds to a magnetic index $m = 1.9$ ($1.5$).}
    \label{Fig:MaxEnergy}
\end{figure}

The top panel of Fig.~\ref{Fig:MaxEnergy} shows the maximum energy $E_\mathrm{p}^\mathrm{max}$ achieved by protons as a function of position $z$, for three models with different sizes of the acceleration region. The behaviour of this function reflects the existence of three regimes in which different loss processes dominate: near the jet base ($z \lesssim 10^9\,\mathrm{cm}$) $E_\mathrm{p}^\mathrm{max}$ increases with $z$ because synchrotron losses dominate at the higher proton energies while the magnetic field drops with $z$. At further distances the energy loss is ruled mainly by adiabatic work, causing $E_\mathrm{p}^\mathrm{max}$ to decrease with $z$ with a constant slope ($z \approx 10^9\,\mathrm{cm}$ to $z \approx 10^{14}\,\mathrm{cm}$). Finally, near the top of the acceleration region, the escape rate overcomes the adiabatic losses and limits $E_\mathrm{p}^\mathrm{max}$, steepening its decrease ($z \gtrsim 10^{14}\,\mathrm{cm}$). The distance $z$ at which the slope breaks increases with $z_\mathrm{t}$, because escape depends on the size of the emission region (and  $z_{\mathrm{max}}=z_{\mathrm{t}}$ in this discussion). More compact regions have higher escape rates, resulting in lower maximum proton energies. Based on this behaviour, we expect the neutron spectrum to be very steep and dominated by low-energy neutrons produced far from the jet base in the general case, reaching high energies only for compact acceleration regions.

\begin{figure}[h]
    \centering
    \includegraphics[width=0.99\hsize]{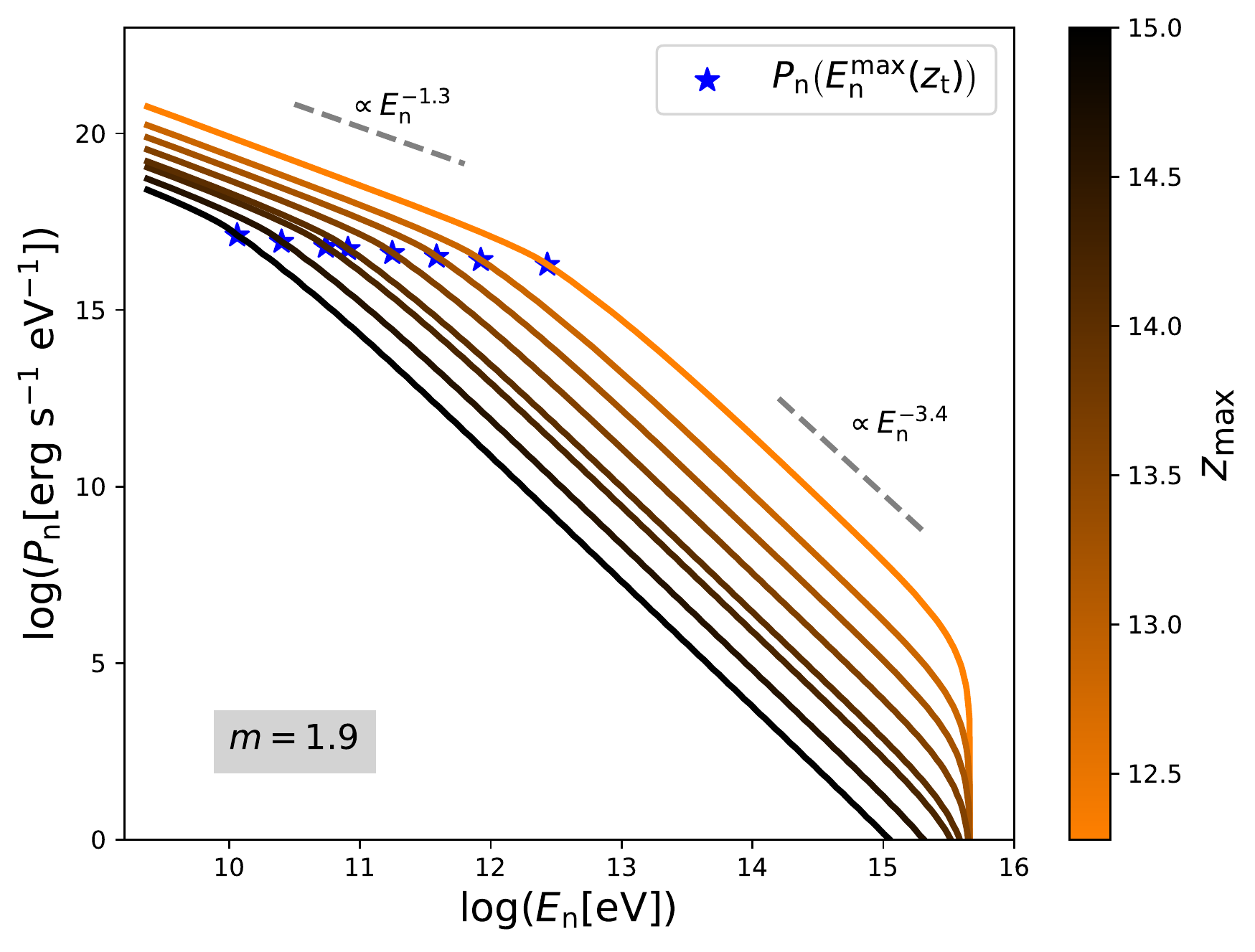}
    \includegraphics[width=0.99\hsize]{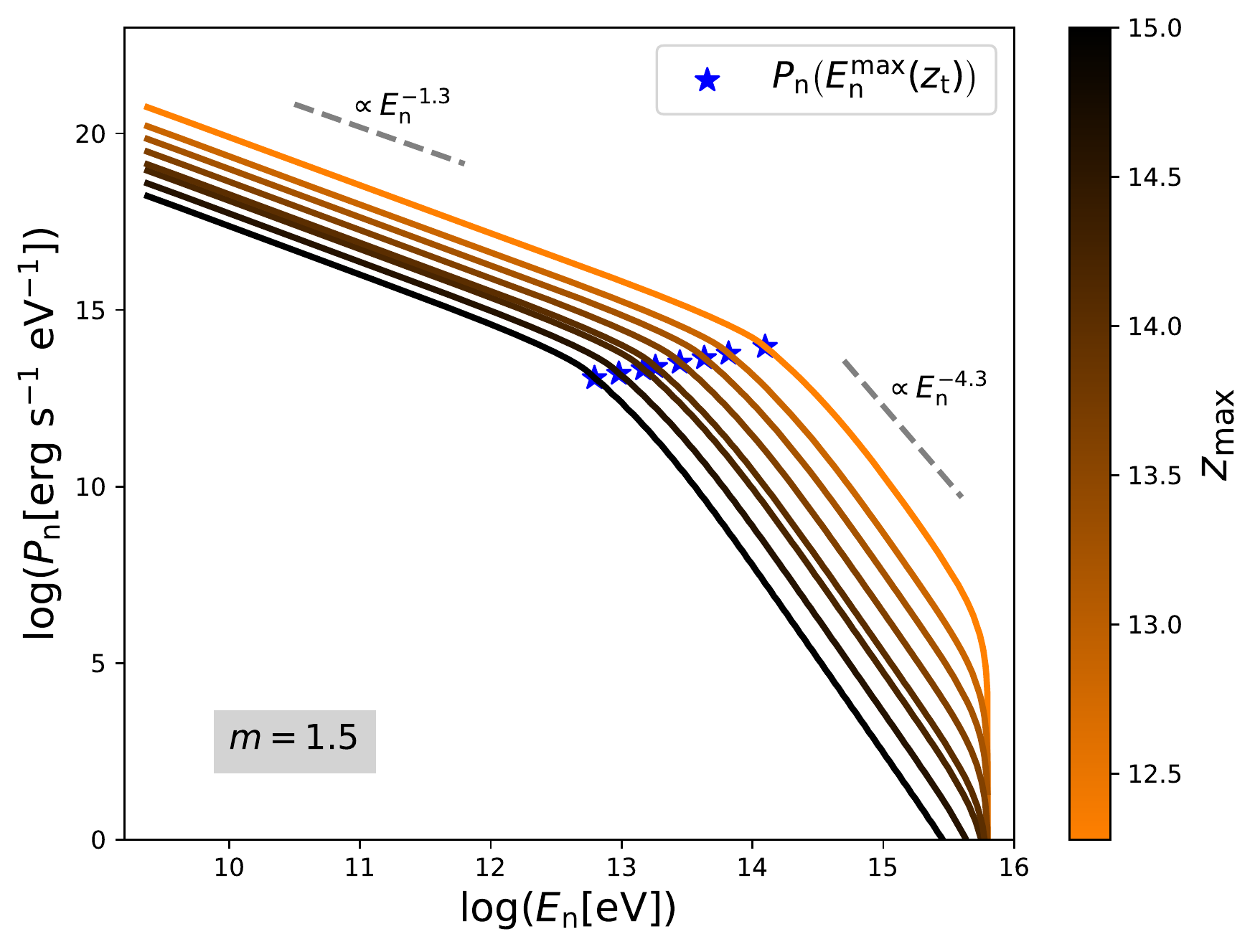}
    \caption{Integrated neutron power spectra in the jet frame, for different values of the top of the acceleration/emission region, $z_{\mathrm{t}} = z_{\mathrm{max}}$. Blue stars indicate the value of $P_{\mathrm{n}}$ evaluated at the maximum neutron energy at $z_{\mathrm{t}}=z_{\mathrm{max}}$, i.e. the minimum value of $E_\mathrm{n}^\mathrm{max} (= 0.5 E_{\mathrm{p}}^{\mathrm{max}})$ along the jet for each model (cf. Fig. \ref{Fig:MaxEnergy}). In all the cases $E_{\mathrm{min}} = 5mc^{2}$. The top (bottom) panel corresponds to a magnetic index $m = 1.9$ ($1.5$). The remaining parameters are those of Table \ref{Tab:modelparameters}.}
    \label{Fig:ZtVariation}
\end{figure}

The top panel of Fig.~\ref{Fig:ZtVariation} shows the integrated neutron spectra of the jet for different values of the top of acceleration region, $z_{\mathrm{t}}$ ($=z_{\mathrm{max}}$ in this scenario). In this case we adopt $E_{\mathrm{min}} = 5\,m_{\mathrm{p}}c^{2}$. Neutron spectra have common features independently of the extension of the emission region: a broken power-law shape with similar spectral indices among all the different scenarios. An index of $3.4$ indicating a very steep behaviour at high energies as predicted by our previous discussion, and a shallower one ($1.3$) at low energies describe extremely well all curves. These features are remarkably independent of the characteristics of the scenario. The break in the spectrum is related to the minimum value of $E_\mathrm{p}^\mathrm{max}$ in each model, which is also $E_\mathrm{p}^\mathrm{max}(z_t)$ and therefore indicates the shrinking of the effective emission region. Neutrons of lower energies are produced in the whole emission region, whereas those of higher energies cannot be produced in the external parts having lower $E_\mathrm{p}^\mathrm{max}$. Below the break, the power emitted in neutrons decreases with energy only due to the corresponding change in acceleration and loss mechanisms. The low-energy index is indeed directly related to the power-law slope of the injected proton energy spectrum due to the weak dependence of losses on the proton energy ($1.3 \approx 1.4 = p - 1$, the decrement in one arising from a factor $E_p$ between the number and the energy of protons). Above the break the change in the effective emission volume produces a faster decrease. The high-energy index is therefore determined by the interplay between the change with $z$ of the maximum proton energy and that of the cross section of the jet.

The above reasoning implies that the value of the high-energy index of the neutron spectrum is ultimately governed by the structure of the jet magnetic field, that determines the variation of $E_\mathrm{p}^\mathrm{max}$. The bottom panel of Fig.~\ref{Fig:MaxEnergy} shows the maximum energy of protons along the jet for a different value of the magnetic index, $m = 1.5$. A strong difference is seen with respect to the the case of $m = 1.9$. A weaker decrease of $B$ with $z$, expressed by a lower value of $m$, translates into a correspondingly weaker decrease of $E_\mathrm{p}^\mathrm{max}$. This enlarges the energy range at which the full jet volume emits, moving the spectral break to higher neutron energies (bottom panel of Fig.~\ref{Fig:ZtVariation}). Below the break, the spectral index remains unchanged for different values of the magnetic index, as the former depends mainly on the injection index $p$. Since the total proton luminosity is fixed, the enlargement of the full-volume spectral range leaves less power available above the break, steepening the high-energy part of the spectrum.

\begin{figure}
    \centering
    \includegraphics[width=0.95\hsize]{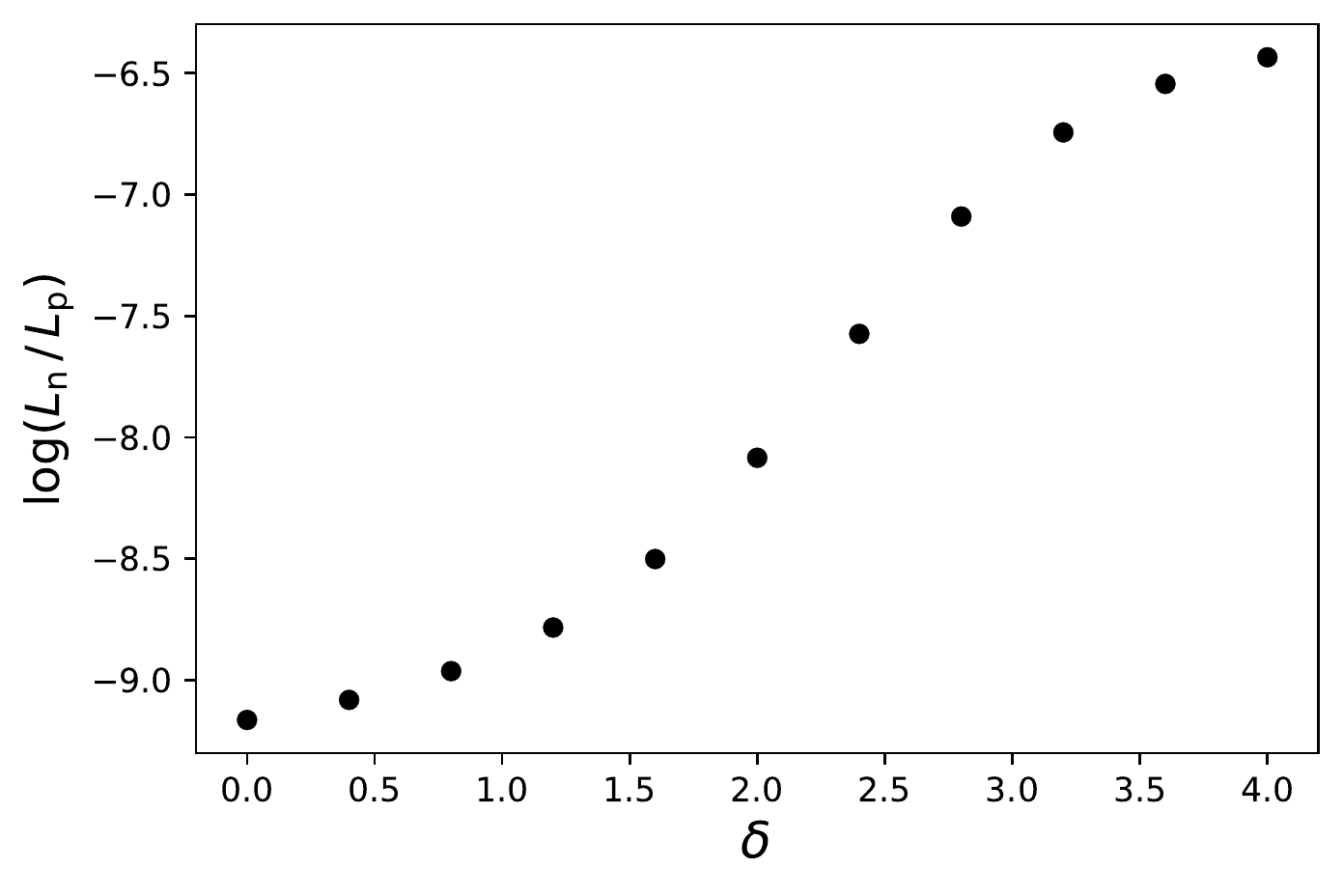}
    \caption{Fraction of the luminosity of relativistic protons transferred to neutrons, as a function of the convective strength of the jet, defined by the parameter $\delta$.}
    \label{Fig:efficiencyvsdelta}
\end{figure}

We also study the effect of different convection regimes on neutron production. For this purpose we use models with a compact acceleration region with $z_\mathrm{t} = 1.9 \times 10^{12}\,\mathrm{cm}$ and an emission region extending to $z_\mathrm{max} = 10^{15}\,\mathrm{cm}$. Convection transports relativistic particles from the former to the latter. As convection is stronger, more power in relativistic protons is transported to regions far from the base, where the density of bulk protons is lower and therefore $p$-$p$ collisions are less frequent. Thus, we expect stronger convective jets to be less efficient producing neutrons. In Fig. \ref{Fig:efficiencyvsdelta} we show the total power of produced neutrons, normalised to the injected proton power, for different convection regimes. In fact, the plot shows that a weakly convective behaviour is more efficient in producing neutrons that a strong one, the power released in neutrons being $\sim 3$ orders of magnitude higher in the former case.

\begin{figure}[h]
    \centering
    \includegraphics[width=0.99\hsize]{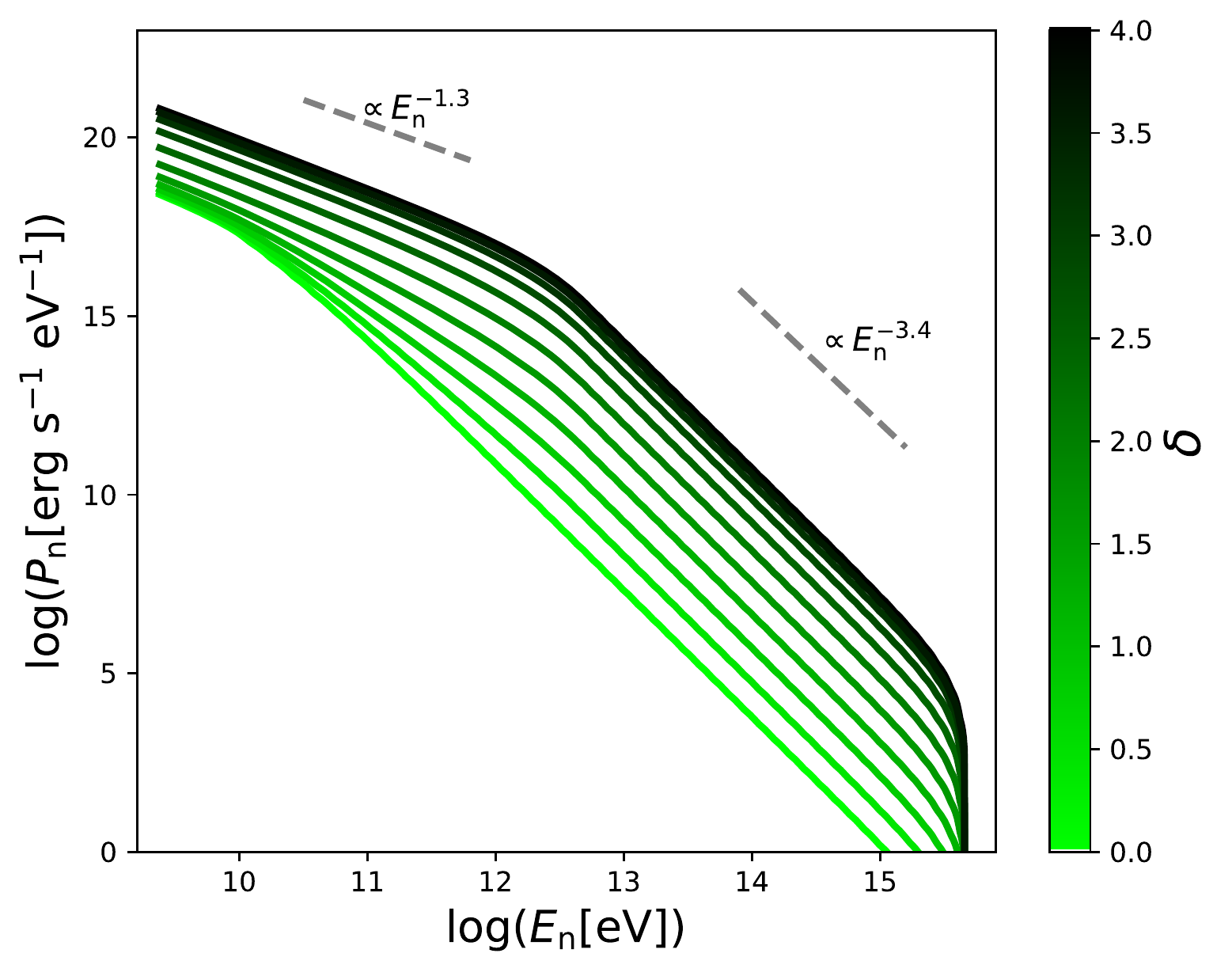}
    \caption{Integrated neutron power spectra in the jet frame, for scenarios with different convective strength, with $E_{\mathrm{min}} = 5mc^{2}$. The remaining parameters are those of Table \ref{Tab:modelparameters}.}
    \label{Fig:ZindexVariation}
\end{figure}

In Fig. \ref{Fig:ZindexVariation} we show the neutron spectra comprising the different convection scenarios. As in the previous case, we use $E_\mathrm{min} = 5\, m_{\mathrm{p}}c^{2}$ (and with the remaining properties of the system as in the fiducial case). Comparing with the fiducial scenario in Fig.~\ref{Fig:ZtVariation}, we see that at values of $\delta \approx 4$ convection can be considered negligible. All neutron spectra show the same common features found in Fig.~\ref{Fig:ZtVariation}: a broken power-law shape with similar spectral indices among the different models. The spectral indices are independent of the convection strength, and their values agree with those of the previous scenario with no convection. The power law breaks at higher energies in the case of a weaker convection regime, which is consistent with our previous finding, because a weaker convection implies a smaller effective emission region.

\subsection{Collimation}
\label{Subsect:collimation}

We now analyse the effect of jet collimation in the production of neutrons. We have already mentioned that the neutron power results mainly from the competition between $p$-$p$ collisions and adiabatic losses. The latter depend strongly on the jet collimation, and if they are reduced, it is expected that proton-proton collisions dominate over a wide range of distances near the base of the acceleration region, increasing the power transferred to neutrons. Escape provides another, but less relevant, competing process. We note from Fig. \ref{Fig:lossrates} that the rates of these three processes are roughly constant along the whole range of proton energies. Thus, such losses are well represented by an average rate over energies, which we denote by $\bar{t}^{-1}$. In order to measure the relevance of proton-proton collisions with respect to other losses for different degrees of jet collimation, we compute the ratio

\begin{eqnarray}
f = \frac{\bar{t}^{-1}_{\mathrm{pp}}}{\bar{t}_{\mathrm{ad}}^{-1} + \bar{t}^{-1}_{\mathrm{esc}}}.
\end{eqnarray}

\begin{figure}
    \centering
    \includegraphics[width=0.99\hsize]{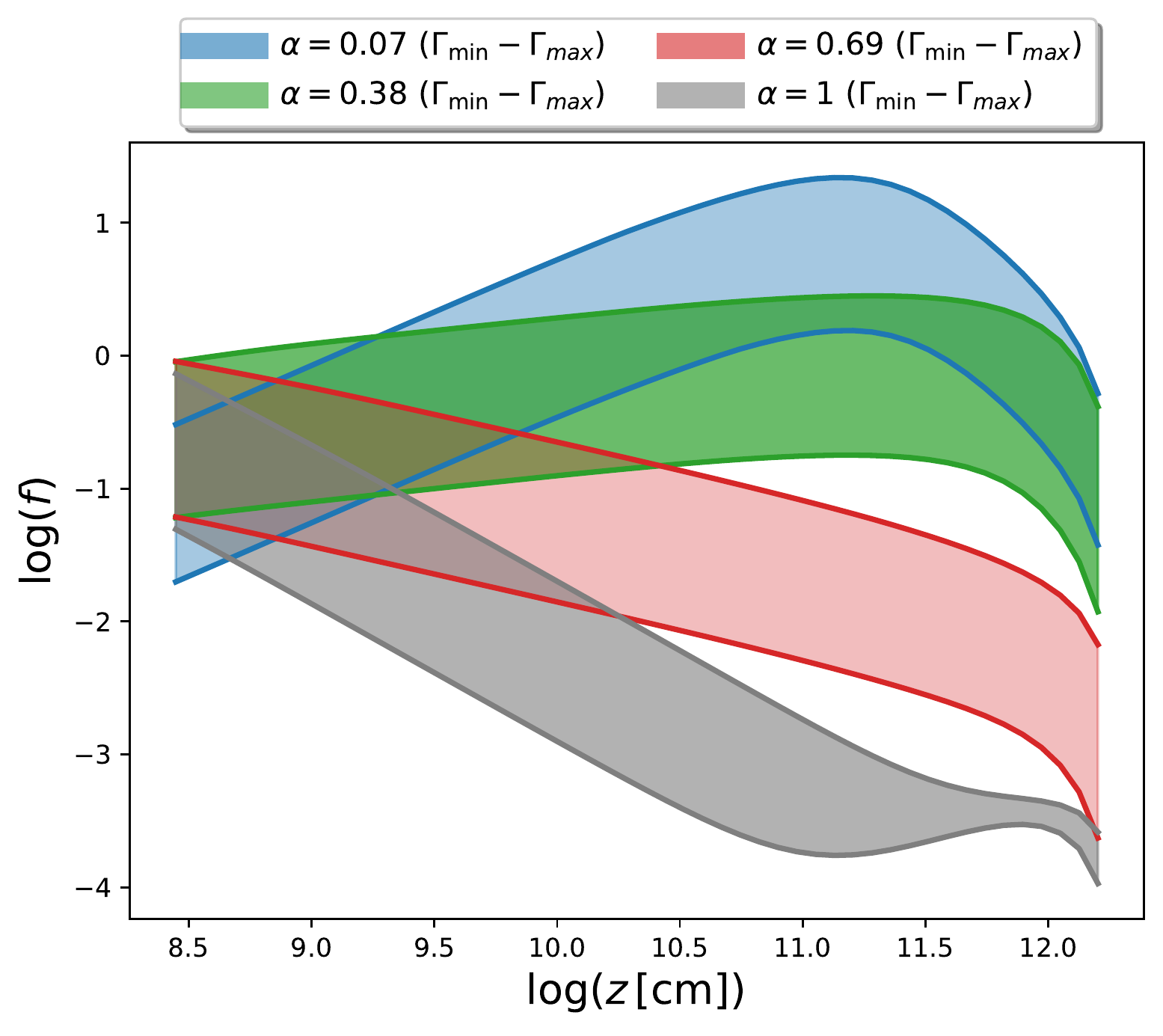}
    \caption{Ratio of proton$-$proton to adiabatic-plus-escape average loss rates, $f$, for models with different collimation parameter $\alpha$ and within a typical range of bulk Lorentz factors, $\Gamma_{\mathrm{min}} \leq \Gamma \leq \Gamma_{\mathrm{max}}$, where $\Gamma_{\mathrm{min}} = 1.034 $ and $\Gamma_{\mathrm{max}} = 5$.}
    \label{Fig:adiabatic}
\end{figure}

\begin{figure}
    \centering
    \includegraphics[width=0.95\hsize]{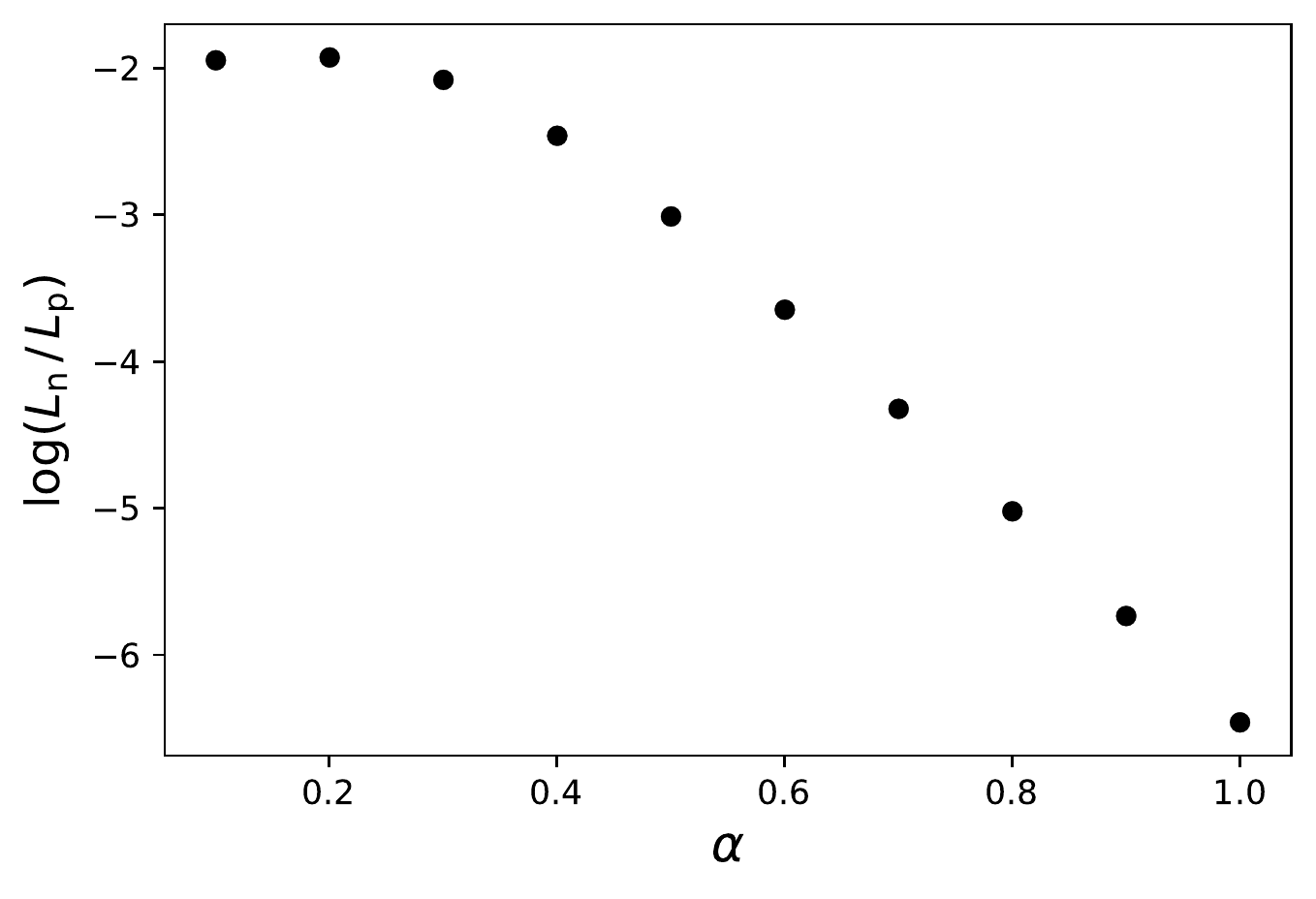}
    \caption{Fraction of the luminosity of protons transfered to neutrons for scenarios comprising different degrees of collimation of the jet.}
    \label{Fig:efficiencyvsalpha}
\end{figure}

\noindent Synchrotron losses are not taken into account since they are not relevant for this analysis, because they only contribute very near the jet base at the highest energies. 
Fig. \ref{Fig:adiabatic} shows $f$ as a function of the distance along the jet, $z$, for the scenario where the jet varies from low ($\alpha = 1$) to high collimation ($\alpha \rightarrow 0$) through the shape parameter $\alpha$. Also, different bulk Lorentz factors are explored, as they also determine the level of adiabatic losses. The range is the same as in \citet{Escobar2021}, $\Gamma_{\mathrm{min}} = 1.034$ \citep[e.g. that of the jet in the microquasar \object{SS 433};][]{Chaty2007} and $\Gamma_{\mathrm{max}} = 5$ The remaining parameters are those of the fiducial case. It can be seen that independently of the value of $\alpha$, higher $p$-$p$ loss rates are obtained if $\Gamma$ is lower. This is in accordance with results of the previous work \citep{Escobar2021}. The shape of the jet has also a major impact on the energetic loss balance. In all cases the $p$-$p$ loss rate is more relevant at regions closer to the jet base, where matter target fields are denser. Well collimated jets reduce adiabatic losses, redirecting energy to the $p$-$p$ channel. In the case $\alpha = 0.07$ (an almost cylindrical jet), proton-proton collisions dominate over the rest of the losses for almost all values of $\Gamma$ and over the whole region of acceleration. This suggests that highly collimated jets are efficient neutron production sites, which is confirmed by Fig.~\ref{Fig:efficiencyvsalpha}. We observe that in the case of a parabolic jet, the total power in the neutron population rises $\sim 4$ orders of magnitude with respect to the conic case.

\begin{figure}
    \centering
    \includegraphics[width=0.99\hsize]{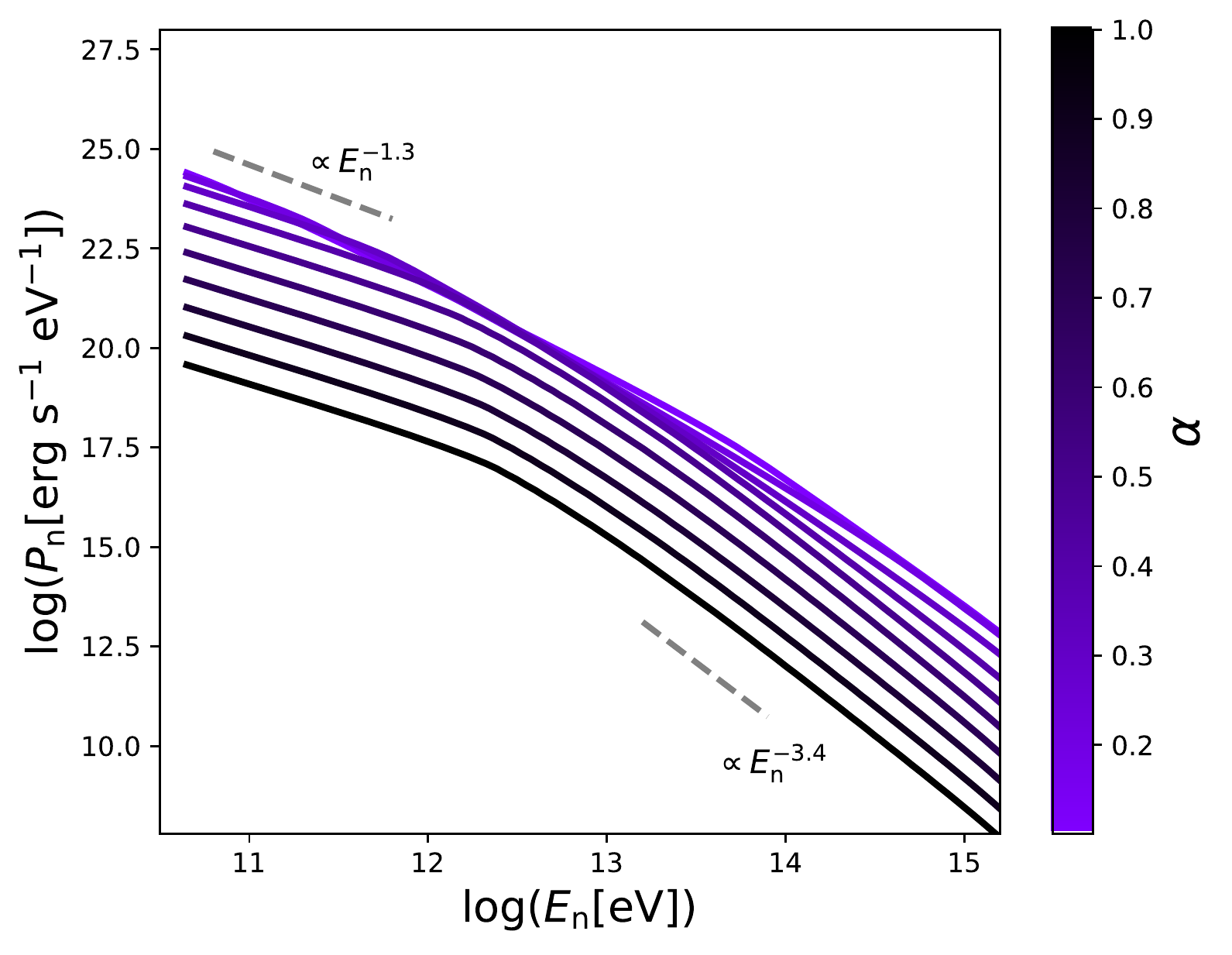}
    \caption{Neutron power spectra in the jet frame, for different values of the collimation parameter $\alpha$ and $E_{\mathrm{min}}=95.4mc^{2}$. The remaining parameters are those of Table \ref{Tab:modelparameters}.}
    \label{Fig:CollimationVar}
\end{figure}

In Fig. \ref{Fig:CollimationVar} we show the integrated spectrum of relativistic neutrons for models comprising different degrees of collimation of the jet. Again, all the spectra are broken power laws with similar spectral indices independently of the collimation regime. Moreover, the spectral indices agree well with those obtained in all previous scenarios. The energy at which the spectra break display smaller changes in this case, shifting to higher values as the jet is more collimated. This is consistent with our previous interpretation because more collimated jets show smaller adiabatic losses and therefore higher $E_\mathrm{p}^\mathrm{max}$ at the same $z$.

From the previous analysis we conclude that slow and collimated jets, with a compact region of acceleration close to the base and with negligible convection, may be very efficient sources of relativistic neutrons and therefore, of cosmic rays.

\subsection{Photohadronic interactions}
\label{Sec:photohadronic}

\begin{figure}
    \centering
    \includegraphics[width=0.99\hsize]{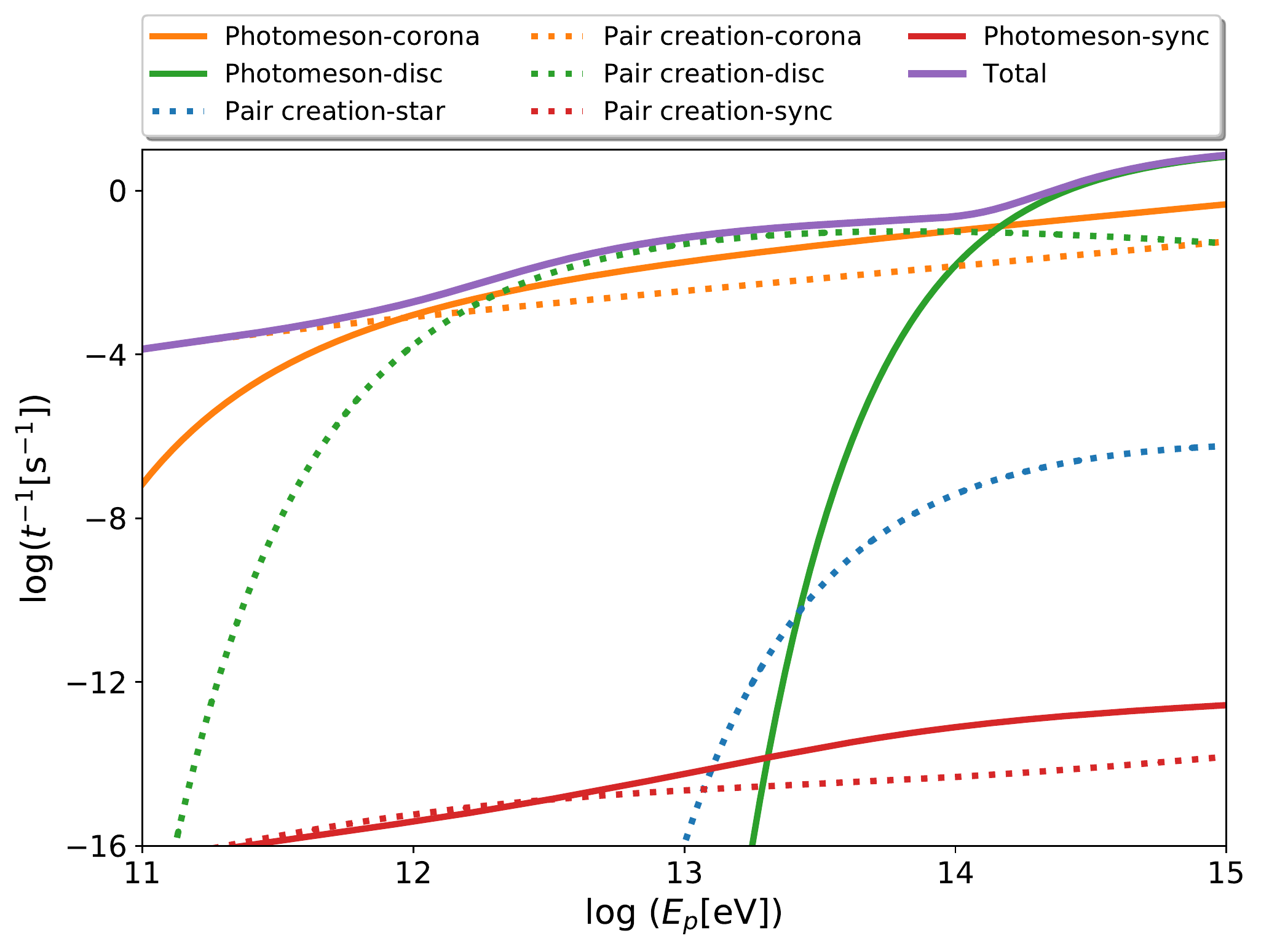}
    \caption{Loss rates for photohadronic interactions between protons and the photon fields of the companion star (blue), the accretion disc (green), the corona (orange) and local synchrotron emission in the jet (red). Dotted lines show the loss rate for the pair-creation channel and solid lines that of the photomeson channel. The total loss rate of the interaction (including pair-creation and photomeson channels) is plotted in solid puprle line.}
    \label{Fig:photohadronic}
\end{figure}

The interaction of relativistic protons with radiation fields also provides a mechanism to produce relativistic neutrons, through some of its branching channels, as discussed in Sec.~\ref{Sec:JetModel}. In this section we assess the contribution of this interaction. We consider all the relevant radiation fields in a MQ, both internal and external to the jet. Given that in the jet frame the relativistic protons have an isotropic momentum distribution, we assume that the photomeson interactions are also isotropic, regardless of the direction of the target photons. In the first case, we adopt the synchrotron photon field radiated from both relativistic electrons and protons, in the local approximation of \citet{Ghisellini1985}. In this case, the photon density can be computed as

\begin{eqnarray}
n_{\mathrm{sy}}^{\mathrm{(i)}} = \frac{q_{\mathrm{sy}}^{\mathrm{(i)}}\,r(z)}{c\,\varepsilon_{\mathrm{ph}}},
\end{eqnarray}

\noindent where $i = \mathrm{e}, \mathrm{p}$, for electrons and protons, respectively, $\varepsilon_{\mathrm{ph}}$ is the photon energy, and $q_{\mathrm{sy}}^{\mathrm{(i)}}$ is the synchrotron emissivity ($[q_{\mathrm{sy}}^{\mathrm{(i)}}] = \mathrm{s}^{-1}~\mathrm{cm}^{-3}$). \footnote{Proton synchrotron losses might not achieve the stationary state given that the loss-rate for the process is low in comparison with the rest of the interactions. As we can see from Fig. \ref{Fig:photohadronic}, however, the contribution of this process to the proton cooling is negligible even with the assumption of a steady emission. Thus, in the remaining cases the contribution will be negligible as well.}. Regarding the external fields, we consider those of the companion star, the accretion disc and the corona surrounding the black hole. For the stellar radiation we adopt a black body with effective temperature $T_{\star}\approx 30\,000$~K in the case of the fiducial model (representing that of the O9.7 companion star of the \object{Cygnus X-1} system). The photons emitted by the corona can be described with a power law of the form \citep[e.g.][]{VieyroSestayo2012}

\begin{eqnarray}
n_\mathrm{c}(\varepsilon_{\mathrm{ph}}) = Q_{\mathrm{c}}\varepsilon_{\mathrm{ph}}^{-\beta}\exp \left( \frac{\varepsilon_{\mathrm{ph}}}{\varepsilon_{\mathrm{ph}}^{\mathrm{max}}} \right),
\end{eqnarray}

\noindent where $\varepsilon_{\mathrm{ph}}^{\mathrm{max}} \approx 150$~KeV. The field is normalised to the total power of the corona, assumed to be $1 \%$ of the Eddington luminosity of the black hole. For the black-hole mass we adopt a value of $M_{\mathrm{BH}} \approx 21~M_{\sun}$, which corresponds to that of the Cygnus X-1 system \citep[][]{MillerJones2021}. Finally, the disc emission is computed adopting a geometrically thin, optically thick accretion disc. The temperature profile is given by \citep[][]{RomeroVila2014} 

\begin{eqnarray}
T(R) = \left[ \frac{3GM\dot{M}}{8\pi\sigma_{\mathrm{SB}}R^{3}} \left( 1 - \sqrt{\frac{R_{\mathrm{in}}}{R}} \right)\right]^{\frac{1}{4}},
\end{eqnarray}

\noindent where $R$ is the radial coordinate of the disc, $\sigma_{\mathrm{SB}}$ is the Stephan-Boltzmann constant and $\dot{M}$ is the accretion rate. The disc extends from $R_{\mathrm{in}} = 0.9 R_{\mathrm{c}}$ to $R_{\mathrm{out}} = 100 R_{\mathrm{in}}$, where $R_{\mathrm{c}} = 35 R_{\mathrm{g}}$ is the radius of the corona and $R_{\mathrm{g}}$ the gravitational radius.

\begin{figure*}[ht]
    \centering
    \includegraphics[width=0.95\hsize]{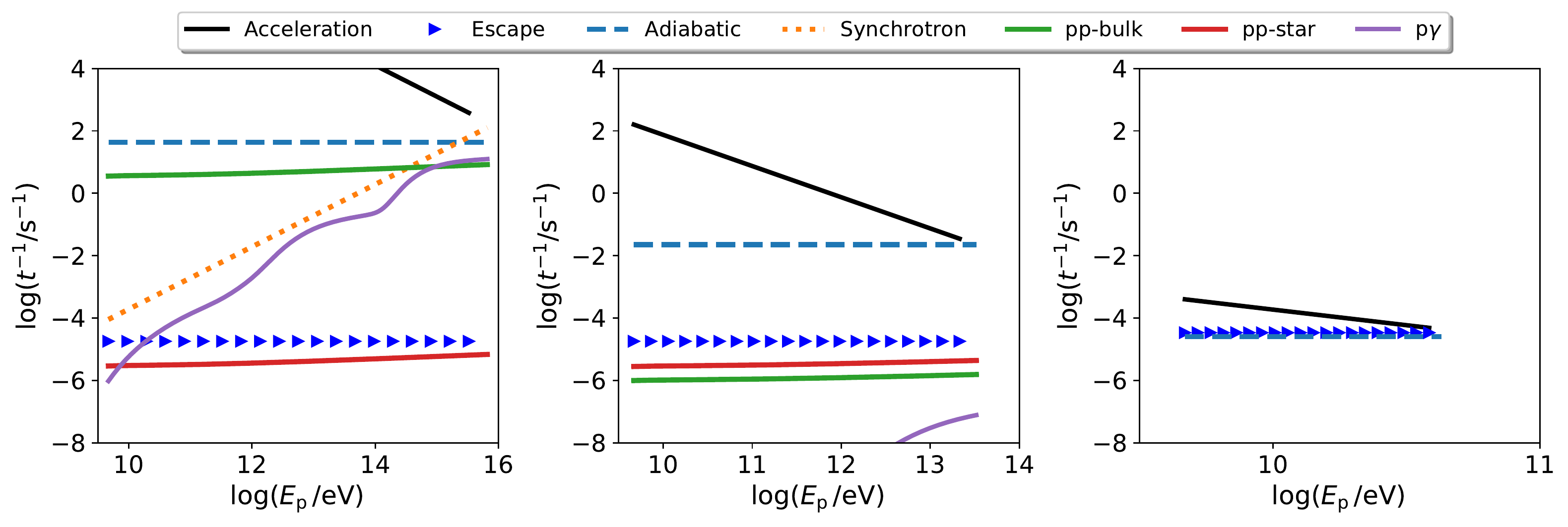}
    \caption{Proton cooling rates including photohadronic interactions at the base (left panel), logarithmic midpoint (middle panel), and the end of the acceleration region (right panel). For this analysis we use the same parameters as in Fig. \ref{Fig:lossrates}.}
    \label{Fig:lossrates_photohadronic}
\end{figure*}

Fig. \ref{Fig:photohadronic} shows the loss rates for photohadronic interactions between protons and the relevant photon fields, at the base of acceleration region. In Fig. \ref{Fig:lossrates_photohadronic} we show the total loss rate for these interactions together with the rest of the losses for protons at the base, logarithmic midpoint, and top of the acceleration region. It is clear that the rate of photohadronic interactions falls well below that of proton-proton collisions, except near the maximum energy of protons at the base of the jet, where disc photons provide a rate similar to that of $p$-$p$. Near $E_{\mathrm{p}}^{\mathrm{max}}$, however, the amount of power released in neutrons is negligible because synchrotron and adiabatic losses dominate, and the proton population carries a low fraction of the total energy of the population (taking into account typical values of the injection spectral index, $1.5 \lesssim p \lesssim 2.4)$. As long as we consider regions further above the base of the jet, photohadronic losses become less relevant since the maximum energy of protons decreases (see also Fig. \ref{Fig:MaxEnergy}). Therefore, the photohadronic contribution to neutron production is negligible in every case.

\subsection{Cosmic-ray spectrum}

In this section we compute the spectra of CRs produced as consequence of neutron decay far from the jet. Given the neutron energies in our model, the fraction of them that decay inside the jet is negligible. We focus on the secondary protons product of neutron beta decay. Electrons and neutrinos are also emitted in the process but they carry a negligible fraction of the progenitor neutron energy ($\sim 0.1\%$). We compute the distribution of neutron decay distances using a probability exponential law in distance as in \citet{Escobar2021}, with mean equal to the lifetime of the neutron times its  velocity, both in the ISM reference frame.

We compute the propagation of protons until they reach the ISM, assuming that they undergo elastic scattering with magnetic plasma fluctuations of the companion wind. We adopt as the boundary between the system and the ISM the sphere where the wind and external medium pressures become equal. We use the formulae given by \citet[][]{Gleeson1978} and \citet{Strauss2011} for the same process in the solar wind, which in our case yields 

\begin{eqnarray}
	\label{eq:lorentz-variation}
    \dot{\gamma} = -\frac{2}{3}\gamma \beta^{2} \frac{v_{\mathrm{w}}}{r},
	\end{eqnarray}
	
\noindent for the energy-loss rate of CR particles. In the previous expression, $\gamma$ is the Lorentz factor of the particles, $v_{\mathrm{w}}$ the stellar wind velocity and $r$ the radial coordinate (centred in the binary system, in this case). These energy losses are the result of particles propagating diffusively in the cavity through scattering off magnetic waves in the plasma. Adopting a diffusive radial motion with Bohm diffusion coefficient \footnote{The Bohm regime corresponds to the case of minimum diffusion time, and in fact is the case in which the spectra could be largely modified. Therefore, other diffusion regimes are expected to have minor impact on the computed CR spectra.}, the Lorentz factor of a proton that escapes to the ISM, $\gamma_{\mathrm{f}}$, which started to propagate with energy $\gamma m_{\mathrm{p}}c^{2}$ at a distance $r_0$ from the system, is

    \begin{eqnarray}
    \label{eq:escape_energy}
    \gamma_{\mathrm{f}} - \gamma + \frac{1}{2}\ln\left( \frac{\gamma_{\mathrm{f}} - 1}{\gamma_{\mathrm{f}} + 1} \right) = - \frac{2}{9} \frac{v_{\mathrm{w}}}{r_0^{2}}\frac{3 e B_{\star} R_{\star}^{3}}{m_{\mathrm{p}} c^{3}},
    \end{eqnarray}
		
 \noindent where $R_{\star}$ and $B_{\star}$ are the star radius and magnetic field at the stellar surface, respectively. This equation allows us to compute the CR spectrum at the system boundary, assuming the number of particles is conserved throughout the propagation process.
 
\begin{figure}
    \centering
    \includegraphics[width=0.98\hsize]{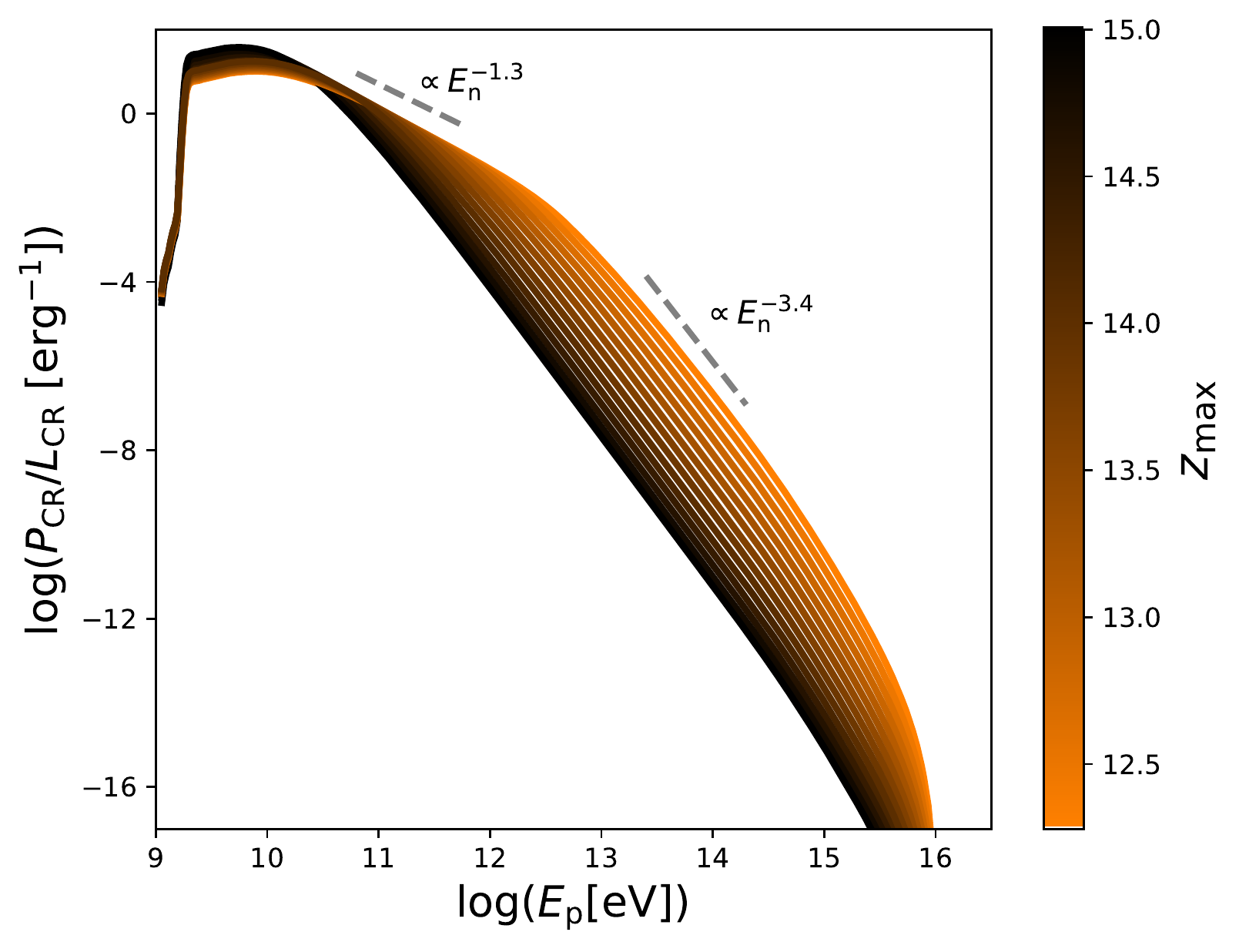}
    \includegraphics[width=0.98\hsize]{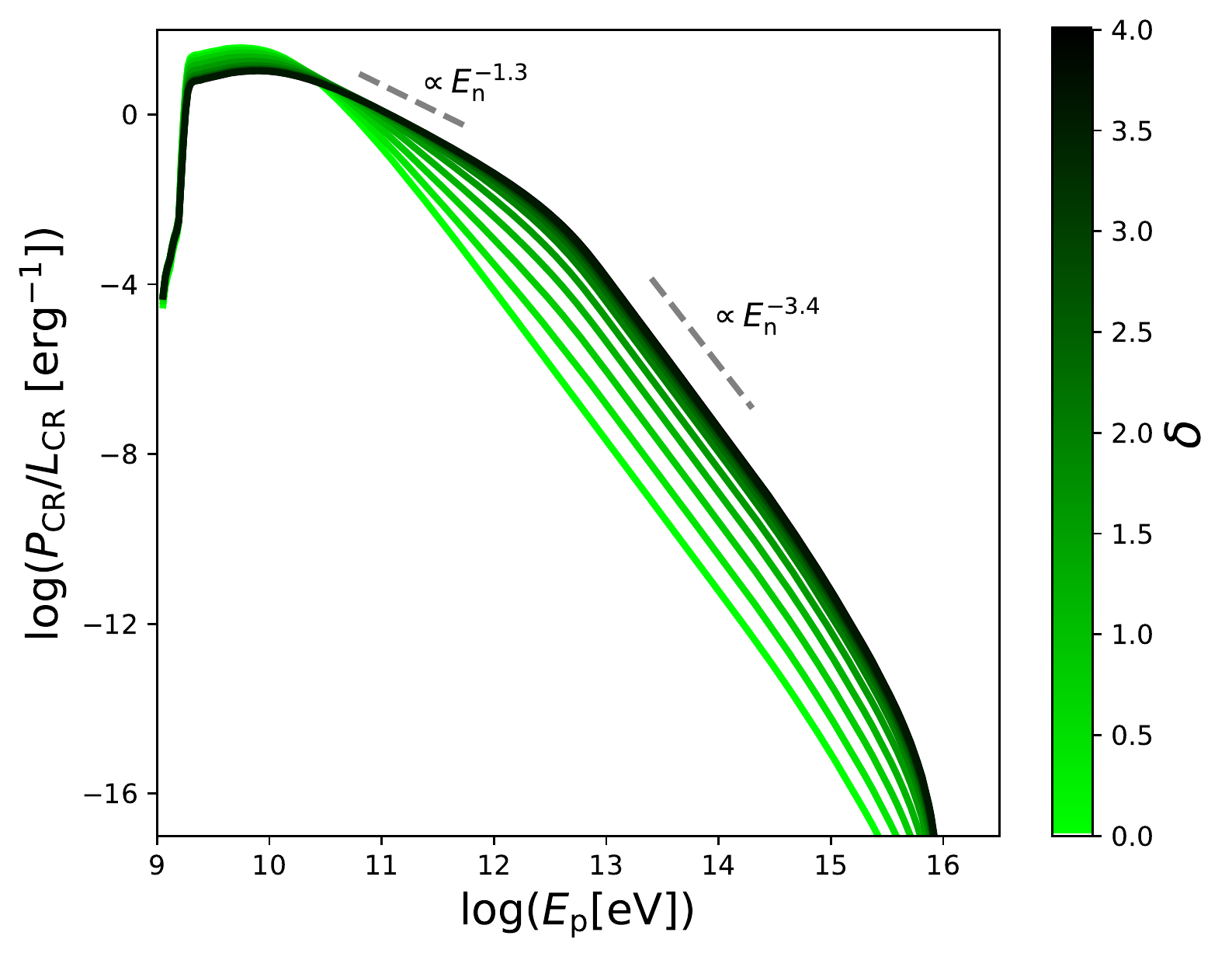}
    \includegraphics[width=0.98\hsize]{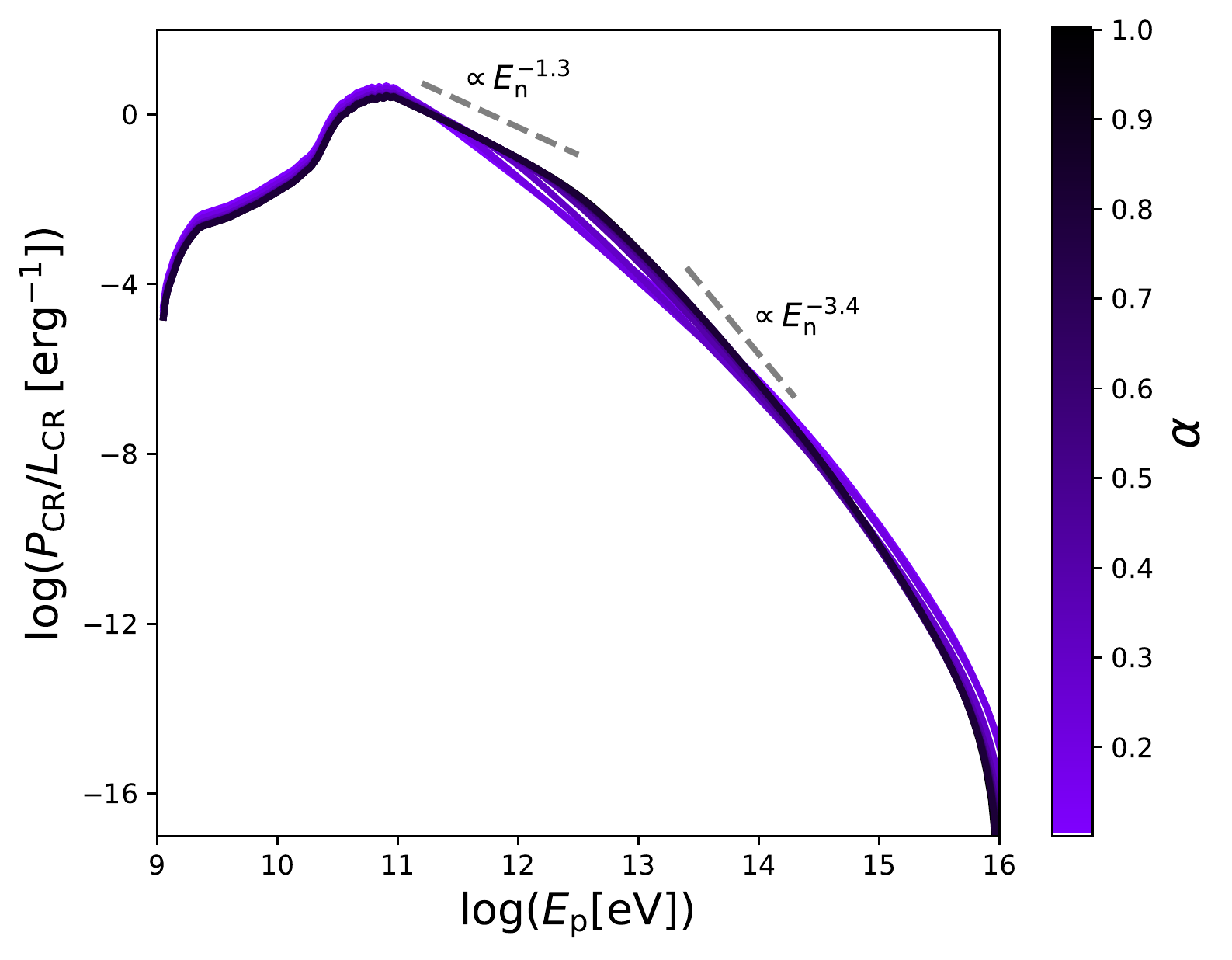}
    \caption{Cosmic-ray spectra for all scenarios discussed in this work, comprising different extensions of acceleration region (top), convection regimes (middle), and degrees of jet collimation (bottom). All spectra are normalised to the total CR luminosity.}
    \label{Fig:crspectra}
\end{figure}

 In Fig.~\ref{Fig:crspectra} we show the normalised CR spectra for all the models discussed in Sec.~\ref{Subsec:accel&conv} and \ref{Subsect:collimation}. In all the cases, at energies $E \gtrsim 10^{11}$~eV, the spectra are essentially the same as those of the neutrons. In addition, CR protons carry essentially all the neutron luminosity produced by the MQ (cf. Fig.~\ref{Fig:efficiencyvsztop}, \ref{Fig:efficiencyvsdelta}, and \ref{Fig:efficiencyvsalpha}). This is because protons suffer almost negligible losses while propagating through the wind plasma. The only consequence of the interaction is the development of a low-energy tail in the spectra, arising from protons cooled below the minimum energy of the parent population. We discuss the implications of these findings in the next section.

\section{Discussion and conclusions}
\label{Sec:Conclusions}

In this work we have explored MQs as sources of CRs, whose production is mediated by relativistic neutrons created in the jet. We have improved upon the work of \citet{Escobar2021} by taking into account the degree of collimation of the jet, convective transport of particles and different sizes of the acceleration and emission regions. We have also included photohadronic interactions with photon fields as another channel that could contribute to produce relativistic neutrons. In this way, we have explored a fairly general scenario for CR emission by a MQ jet, compatible with current observations.

\citet{Escobar2021} have shown that the jet luminosity and Lorentz factor are the main properties affecting the neutron luminosity, and therefore the CR power injected into the ISM. They have estimated the efficiency of the jets as CR sources in the range $\sim 10^{-6}-10^{-4}$. In the present work, we have shown that also the collimation and compactness of the neutron production region, together to its proximity to the jet base, have a major influence on the efficiency of CR production. In the cases explored in this work, the CR power can rise up to $\sim 4$ orders of magnitude for a highly collimated jet with a compact neutron emission region, in comparison with a conic jet. This renders MQ jets a serious candidate to an alternative CR acceleration engine. We note that jets could even develop recollimation shocks that may yield a roughly cylindrical flow. In these cases the efficiency of cosmic-ray production may increase even more, since the density of bulk protons stays nearly constant along the jet and so does the proton-proton collision rate. It is worth to emphasise that the branching ratio of the neutron production channel in $p$-$p$ collisions adopted in this work ($0.16$) is a low, conservative value, and that the rate of neutron and CR production might increase up to a factor of $\sim 6$ if this branching ratio is increased to the level adopted in previous works \citep[][]{Sikora1989,Atoyan1992a, Atoyan1992b, Vila2014, Romero2020}.

We recall that \citet{Pepe2015} model the \object{Cygnus X-1} jet with a conic shape, whereas in a recent model by \citet{Kantzas2021}, the authors assume a well collimated flow with no adiabatic losses. In both works, the authors manage to explain the available data, showing that current observations are still unable to break all model degeneracies. Therefore, for a single source, our model predicts a range of CR luminosities depending on the assumptions about the geometry of the acceleration region. For the case of \object{Cygnus X-1}, \citet{Escobar2021} estimate a CR luminosity of $L_{\mathrm{CR}}\approx 10^{32}~\mathrm{erg}~\mathrm{s}^{-1}$ (including the bulk and wind protons as targets for $p$-$p$ collisions, and the presence of a counterjet). In the present work we have shown that for the same energetics of the relativistic populations, if the jet is assumed to be highly collimated, this amount would increase to $L_{\mathrm{CR}}\approx 10^{35-36}~\mathrm{erg}~\mathrm{s}^{-1}$. On the other hand, a ultraluminous X-ray source (ULX) would provide $\sim 10^{37-38}~\mathrm{erg}~\mathrm{s}^{-1}$ if jets are collimated. The total energy released in the MQ lifetime, adopting a duty cycle of $\sim 10^{6}~\mathrm{yr}$ for a high-mass MQ jet, such as that of \object{Cygnus X-1}, would be $\sim 10^{48-49}~\mathrm{erg}$ in the collimated case, whereas a ULX would directly rival a typical SNR, injecting $10^{50-51}~\mathrm{erg}$. Clearly, an improvement in both multiwavelength observations and theoretical jet models is required to resolve these degeneracies, and estimate the contribution of MQ jets to the CR population with higher precision.

Regarding the proton CR spectrum (see Fig. \ref{Fig:crspectra}), our scenarios produce a broken power law with spectral indices of $\sim 2.3$ (low energy) and $\sim 4.4$ (high energy). The former is remarkably independent of any of the features of the jet (cf. Fig. \ref{Fig:ZtVariation}, \ref{Fig:ZindexVariation} and \ref{Fig:CollimationVar}), and can be traced to the injection index $p$, given that the loss mechanisms that dominate over the whole jet hardly modify the spectral index of the CR outcome. The break marks the maximum neutron energy at the top of the emission region. CRs with energies below the break are the result of the decay of neutrons coming from the full volume of the jet, whereas those above the break come progressively from smaller regions. This shrinking of the emission region produces the steepening of the CR spectrum at high energies. Both the break and the high-energy index show a deep connection with the basic physics of our models, mainly the distribution of magnetic field strength in the jet. Slowly decaying fields along the jet axis increase the energy range at which the full volume emits, moving the break to higher energies and producing a steeper CR spectrum with less power above the break.

The high-energy part of the spectrum is softer than that predicted in the standard SNR scenarios. A harder injection spectral index would produce CR spectra with indices closer to that observed. Moreover, if we compute the propagation of CRs at energies in the range $\sim 10~\mathrm{GeV}- 10~\mathrm{TeV}$ (where the spectral index is $\sim 2.3$) adopting an ISM diffusion coefficient of $D(E)\propto E^{0.3}$  \citep[e.g.][]{Blasi2013}, the spectra would evolve to a power-law with spectral index $\sim 2.6$, very close to that observed \citep[$\sim 2.7$, e.g.][]{Berezinskii1990}, which the current SNR paradigm fails to explain. To fully settle this question, an estimate of the total spectrum of CRs produced by MQs, including the whole population of these sources and propagation effects, is required.

Even though photohadronic interactions are negligible in comparison with other proton losses in MQ jets, neutron production through these channel merits investigation in other scenarios that may provide suitable conditions for the process to take place significantly. This might be the case of MQ coronae, as it was explored in previous works \citep[][]{Vila2014}. These scenarios would not only contribute to the total CR luminosity of MQs, but would also increase their duty cycle, as they would allow CRs to be produced even while a jet is not active. In fact, MQs may undergo phases in which a jet is present for a time, while in the rest of the cycle the system may undergo different accretion episodes forming disc and coronae structures.

Apart from quantitatively establishing MQ jets as serious candidates to compete with SNRs as sources of CRs, our work opens the possibility of making a step forward in a different direction. Our results show that the CR emission depends only on jet properties. External photon fields and winds  (either from the accretion disc, corona or companion star) providing targets for photohadronic and $p$-$p$ interactions, respectively, play a minor role in neutron production. They do not affect the propagation of neutron decay products until they reach the ISM either, except for very small changes at the low energy end of the spectrum. This indicates that only jet parameters are needed to predict the CR injection by a MQ, while the type of companion (i.e. high or low mass) and details of the disc or corona are irrelevant. By quantitatively relating jet parameters such as the Lorentz factor or the jet luminosity with the CR luminosity and spectrum, we pave the way to the construction of models describing the contribution of the whole population of MQs to the Galaxy CRs.

Finally, the mechanism explored in this work provides a tool to investigate the impact of MQs in the reionisation and heating rates in the early Universe. Different works have shown that X-ray binaries, and therefore MQs, would be more numerous and luminous in the Cosmic Dawn \citep[][]{Mirabel2011, Kaaret2011, Brorby2014, Douna2015, Ponnada2019}. Our work provides a way of estimating the contribution of these systems to the CR population, whose role in the modification of the temperature and ionisation state of the intergalactic medium at early epochs is still a matter of debate \citep[][]{Tueros2014, Leite2017, Douna2018}. In this context, our result about the existence of a low-energy tail of CRs produced in the propagation of neutron products in their way to the ISM, is interesting due to the larger ionisation power of low-energy particles \citep[][]{Douna2018}.

\begin{acknowledgements}
GJE and LJP acknowledge the grant PIP 2014/0265 of Argentine National Research Council (CONICET). GJE also acknowledges financial support from the European Research Council for the ERC Consolidator grant DEMOBLACK, under contract no. 770017. GER is supported by the National Agency for
Scientific and Technological Promotion (PICT 2017-
0898 and PICT 2017-2865) and the Spanish Ministerio
de Ciencia e Innovación (MICINN) under grant
PID2019-105510GBC31 and through the Center of Excellence
Mara de Maeztu 2020-2023 award to the ICCUB
(CEX2019-000918-M)
\end{acknowledgements}

   \bibliographystyle{aa} % style aa.bst
   \bibliography{bibliography} % your references Yourfile.bib
\end{document}